\documentclass[10pt,journal,letterpaper]{IEEEtran}

\usepackage{ifpdf}

\ifpdf
\usepackage[pdftex]{graphicx}
\usepackage{mathrsfs}
\usepackage{amsmath}
\usepackage{amsfonts,amssymb}
\usepackage[square, comma, sort&compress, numbers]{natbib}
\usepackage{algorithm}
\usepackage{algorithmicx}
\usepackage[noend]{algpseudocode}

\usepackage{enumerate}
\usepackage{caption}
\usepackage{multirow}
\usepackage{multicol}
\usepackage{array}
\usepackage{tabularx}
\usepackage{makecell}
\usepackage{booktabs}
\usepackage{amsthm}
\usepackage{color}
\usepackage{threeparttable}
\usepackage[square,numbers,sort&compress]{natbib}

\usepackage{lipsum}

\usepackage{array}
\usepackage{subfigure}
\usepackage{subcaption}

\newcommand{\PreserveBackslash}[1]{\let\temp=\\#1\let\\=\temp}
\newcolumntype{C}[1]{>{\PreserveBackslash\centering}p{#1}}
\newcolumntype{R}[1]{>{\PreserveBackslash\raggedleft}p{#1}}
\newcolumntype{L}[1]{>{\PreserveBackslash\raggedright}p{#1}}

\usepackage{marvosym}
\usepackage{CJK}
\usepackage{subfigure}
\usepackage{url}
\usepackage{multirow}
\usepackage{diagbox}
\usepackage{booktabs}

\else
\fi

\ifCLASSINFOpdf
\else
\fi

\begin{document}
	
\title{Across-Platform Detection of Malicious Cryptocurrency Transactions via Account Interaction Learning}
	
\author{
	Zheng Che, 
	Meng Shen,~\IEEEmembership{Member,~IEEE,} 
        Zhehui Tan,
        Hanbiao Du, 
        Wei Wang,~\IEEEmembership{Member,~IEEE,}\\
        Ting Chen,~\IEEEmembership{Member,~IEEE,}
        Qinglin Zhao,~\IEEEmembership{Senior Member,~IEEE,}
        Yong Xie,~\IEEEmembership{Member,~IEEE,}\\
        and Liehuang Zhu,~\IEEEmembership{Senior Member,~IEEE}
	
	\IEEEcompsocitemizethanks{
		\IEEEcompsocthanksitem Z. Che and Z. Tan are with the School of Computer Science, Beijing Institute of Technology, Beijing 100081, China (e-mail: \ chezheng@bit.edu.cn, inksplatter99@163.com).
		\IEEEcompsocthanksitem M. Shen, H. Du and L. Zhu are with the School of Cyberspace Science and Technology, Beijing Institute of Technology, Beijing 100081, China (e-mail: \{shenmeng, duhanbiao, liehuangz\}@bit.edu.cn).
        \IEEEcompsocthanksitem W. Wang is with the Ministry of Education Key Lab for Intelligent Networks and Network Security, Xi’an Jiaotong University, Xi’an 710049, China (e-mail: wangwei1@bjtu.edu.cn).
        \IEEEcompsocthanksitem T. Chen is with the School of Computer Science and Engineering, University of Electronic Science and Technology of China, Chengdu 611731, China (e-mail: brokendragon@uestc.edu.cn).
        \IEEEcompsocthanksitem Q. Zhao is with the School of Computer Science and Engineering, Macau University of Science and Technology, Macau, China (e-mail: qlzhao@must.edu.mo).
        \IEEEcompsocthanksitem Y. Xie is with the School of Computer and Information Science, Qinghai institute of Technology, China, Xining 810016, China (e-mail: mark.y.xie@qq.com).
	}
}
	
\maketitle

\begin{abstract}

With the rapid evolution of Web3.0, cryptocurrency has become a cornerstone of decentralized finance. While these digital assets enable efficient and borderless financial transactions, their pseudonymous nature has also attracted malicious activities such as money laundering, fraud, and other financial crimes. Effective detection of malicious transactions is crucial to maintaining the security and integrity of the Web 3.0 ecosystem. Existing malicious transaction detection methods rely on large amounts of labeled data and suffer from low generalization. Label-efficient and generalizable malicious transaction detection remains a challenging task. In this paper, we propose \emph{ShadowEyes}, a novel malicious transaction detection method. Specifically, We first propose a generalized graph structure named \emph{TxGraph} as a representation of malicious transaction, which captures the interaction features of each malicious account and its neighbors. Then we carefully design a data augmentation method tailored to simulate the evolution of malicious transactions to generate positive pairs. To alleviate account label scarcity, we further design a graph contrastive mechanism, which enables ShadowEyes to learn discriminative features effectively from unlabeled data, thereby enhancing its detection capabilities in real-world scenarios. We conduct extensive experiments using public datasets to evaluate the performance of ShadowEyes. The results demonstrate that it outperforms state-of-the-art (SOTA) methods in four typical scenarios. Specifically, in the zero-shot learning scenario, it can achieve an F1 score of 76.98\% for identifying gambling transactions, surpassing the SOTA method by 12.05\%. In the scenario of across-platform malicious transaction detection, ShadowEyes maintains an F1 score of around 90\%, which is 10\% higher than the SOTA method.

\end{abstract}
               

\IEEEpeerreviewmaketitle

\vspace{25pt}
\section{Introduction}\IEEEPARstart{T}{he} advent of Web 3.0 has brought about a paradigm shift in the digital world, emphasizing decentralization, user sovereignty, and enhanced security. As a crucial part of the Web 3.0 technology framework, cryptocurrency's widespread adoption has facilitated legitimate financial transactions but also attracted malicious actors exploiting its 
pseudonymous nature to engage in illicit activities such as money laundering, fraud, and other financial crimes. According to Chainalysis \cite{ChainalysisNews}, the total amount of malicious funds transferred through money laundering activities reached approximately \$6.7 billion in 2023. Additionally, the U.S. Federal Trade Commission reported that since 2021, consumers have reported over \$1 billion in cryptocurrency scam losses\footnote{https://www.ftc.gov/news-events/news/press-releases/2022/06/new-analysis-finds-consumers-reported-losing-more-1-billion-cryptocurrency-scams-2021}. Therefore, detecting malicious transaction behaviors has become crucial for maintaining the Web 3.0 ecosystem.

\begin{figure}[htbp]
	\small
	\centering
	\includegraphics[width=\linewidth]{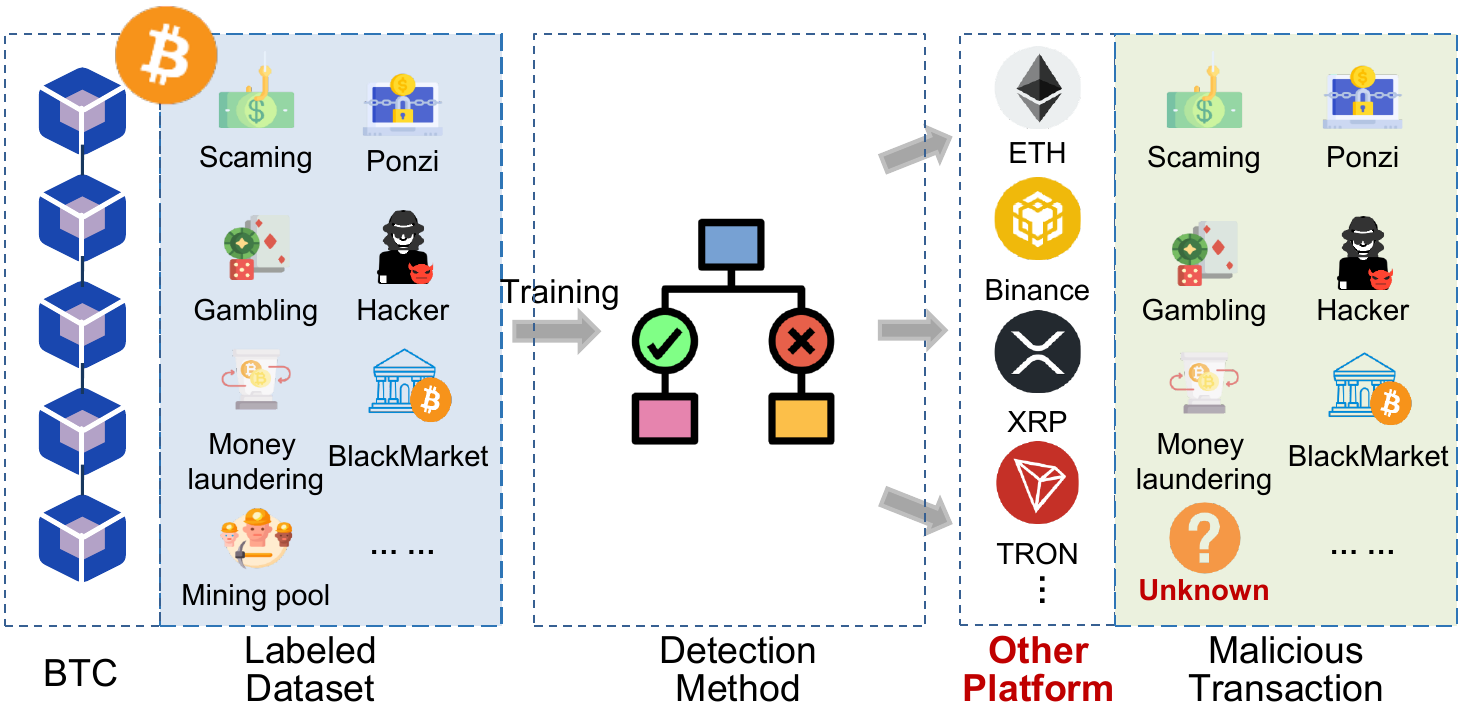}
	\caption{Malicious transaction detection in cryptocurrencies.}
	\label{fig:TxClass}
\end{figure}

An efficient malicious transaction detection method should be able to identify various common and unknown malicious transactions across platforms, as depicted in Fig. \ref{fig:TxClass}. Recent studies have proposed several methods for detecting malicious transactions, which can be categorized into two types based on their technical approaches. The first type involves directly detecting malicious transactions using expert experience or designing hand-crafted features with behavioral discernibility \cite{CCS23}\cite{WWW18}. Although these methods are effective for specific malicious behaviors, they often struggle with generalization, leading to lower accuracy when dealing with complex, diverse, and rapidly evolving malicious activities. The second type introduces interactions among transaction addresses to create representations based on transaction graph features \cite{TTAGN}\cite{JSAC22}. This approach, which considers the collective behavior of malicious transaction accounts, exhibits greater robustness. However, it is heavily dependent on specific network structures, posing challenges in transferring across different datasets and blockchain platforms. Moreover, both types of methods typically rely on labeled data, and the scarcity of labeled datasets in real-world scenarios severely limits their effectiveness.

To tackle these challenges, we propose ShadowEyes, a method for detecting malicious transactions based on graph representation and contrastive learning. To achieve effective detection, we design a data augmentation method to enhance sample diversity by simulating the transaction behaviors of malicious users to evade regulation. To overcome the challenge of poor method portability across platforms with different structures, we design a novel representation called \emph{TxGraph} that leverages account interaction features, integrating common characteristics from diverse blockchain platforms. Furthermore, to mitigate the scarcity of labeled data, we design a training mechanism grounded in graph contrastive learning. This mechanism enables ShadowEyes to learn discriminative features effectively from unlabeled data, thus enhancing its detection capabilities in real-world scenarios.

We summarize the main contributions as follows:

\begin{itemize}

\item We propose a novel transaction representation named \emph{TxGraph} based on account interaction features. \emph{TxGraph} can extract distinct features that differentiate between various malicious transactions and robust transaction features for different cryptocurrency platforms, such as BTC and ETH. Our experiments validate that \emph{TxGraph} is more effective than the two typical representations employed by state-of-the-art methods.
\item We propose ShadowEyes, an effective malicious transaction detection method based on contrastive learning. The encoder in ShadowEyes can be pre-trained with unlabeled transaction samples. In the retraining phase, we only need a few labeled samples to train the transaction classifier to achieve high accuracy in detecting new malicious transactions.
\item We design a data augmentation method that simulates various types of malicious transactions. This approach increases the diversity of pre-training samples, providing ShadowEyes with a richer set of examples to learn from. The enhanced training dataset improves the ShadowEyes's generalization capabilities and robustness in detecting a broad spectrum of malicious activities.
\item We evaluate the performance of ShadowEyes by comparing it to state-of-the-art methods, including FICL \cite{ProCL}, BERT \cite{Bert4ETH}, DT \cite{DesicionTree}, KNN \cite{Arxiv-KNN}, LR \cite{KDD23}, MLP \cite{MLP}, ABGRL \cite{TNSE24}, and CNN \cite{CNN}. ShadowEyes achieves the highest classification F1 scores across multiple scenarios, such as zero-shot learning, across-platform detection, imbalanced datasets, and few-shot learning.
\end{itemize}

The remainder of this paper is organized as follows. We first introduce the background of malicious behavior in cryptocurrency and summarize the related work in Section \ref{sec:Related Work}. Then, we describe the design goals in Section \ref{subsec:Designgoal}. We introduce the transaction representation in Section \ref{sec:representation} and present ShadowEyes in Section \ref{sec:ShadowEyes}. Next, we evaluate the performance of ShadowEyes and compare it comprehensively with the state-of-the-art methods in Section \ref{sec:Experiments}. Finally, we discuss the limitations of ShadowEyes and future research directions in Section \ref{sec:discussion} and we conclude this paper in Section \ref{sec:Conclusion}.

\begin{table*}[!t]  
    \footnotesize  
    \centering  
    \caption{The comparison with the existing malicious transaction detection methods.}  
    \label{tab:related work}  
    \begin{threeparttable}
    \renewcommand{\arraystretch}{1.1}
    \begin{tabularx}{\textwidth}{>{\centering\arraybackslash}p{2.5cm}|c|c|c|c|*{5}{>{\centering\arraybackslash}X}} 
    \toprule  
    \multirow{2}{*}{Categories} & \multirow{2}{*}{Typical Methods} & \multirow{2}{*}{Transaction Representation} & \multirow{2}{*}{Classifier} & \multirow{2}{*}{Portability} & \multicolumn{5}{c}{Detecting Transaction\textsuperscript{1}} \\  
    \cline{6-10}  
     &  &  &  & & Phi. & ML. & Pon. & Oth. & \textbf{Unk.} \\  
    \midrule  
    \multirow{4}{2.2cm}{Empirical Analysis}  
     & MFScope \cite{MFScope} & Transaction features & Expert judgment &\textcolor{red}{\texttimes} & & & &\textcolor{blue}{\checkmark} & \\
     & TxPhishScope \cite{CCS23} & Transaction features & Expert judgment &\textcolor{red}{\texttimes} &\textcolor{blue}{\checkmark} & & & & \\  
     & PGDetector \cite{TIFS24Fishing} & Transaction and motif features & PageRank &\textcolor{red}{\texttimes} &\textcolor{blue}{\checkmark} & & & & \\
     & XBlockFlow \cite{TIFS24ML} & Transaction features & Community detection &\textcolor{red}{\texttimes} & &\textcolor{blue}{\checkmark} & & & \\ 
    \cline{1-10}  
    \multirow{5}{2.2cm}{Machine Learning}  
     & Elmougy et al. \cite{KDD23} & Transaction features & LR &\textcolor{red}{\texttimes} & &\textcolor{blue}{\checkmark} & &\textcolor{blue}{\checkmark} & \\  
     & Farrugia et al. \cite{ESWA20} & Statistical features & XGBoost &\textcolor{red}{\texttimes} & & & &\textcolor{blue}{\checkmark} & \\  
     & Chen et al. \cite{WWW18} & Accounts and operation codes & XGBoost &\textcolor{red}{\texttimes} & & &\textcolor{blue}{\checkmark}& & \\
     & MulCas. \cite{MulCas} & Bytecode and semantic features & Cascade ensemble &\textcolor{red}{\texttimes} & & &\textcolor{blue}{\checkmark}& & \\
     &Bert4ETH \cite{Bert4ETH} & Transaction features & BERT &\textcolor{red}{\texttimes}  &\textcolor{blue}{\checkmark} & & &\textcolor{blue}{\checkmark} & \\
    \cline{1-10}
    \multirow{4}{2.2cm}{Graph Learning}  
     & Jin et al. \cite{JSAC22} & Heterogeneous features & XGBoost &\textcolor{red}{\texttimes} & & & &\textcolor{blue}{\checkmark} & \\  
     & TTAGN \cite{TTAGN} & Heterogeneous features & LightGBM &\textcolor{red}{\texttimes} &\textcolor{blue}{\checkmark} & & & & \\  
     & ABGRL \cite{TNSE24} & Graph features & RF &\textcolor{red}{\texttimes} &\textcolor{blue}{\checkmark} & & & & \\
\cline{2-10}
     &\textbf{ShadowEyes} & Heterogeneous features & MLP & \textcolor{blue}{\checkmark} & \textcolor{blue}{\checkmark} & \textcolor{blue}{\checkmark} & \textcolor{blue}{\checkmark} & \textcolor{blue}{\checkmark} & \textcolor{blue}{\checkmark} \\  
    \bottomrule  
    \end{tabularx}
    \begin{tablenotes}
      \item[1] In the \emph{Detecting Transaction} columns, \emph{Phi.} is phishing fraud transaction, \emph{ML.} is money laundering transaction, \emph{Pon.} is Ponzi scheme transaction, \emph{Oth.} is other type of malicious transaction, \emph{Unk.} is unknown malicious transaction.
    \end{tablenotes}
    \end{threeparttable}   
    \vspace{-0.2cm}
\end{table*}

\section{Background And Related Work}\label{sec:Related Work}
In this section, we first introduce existing types of malicious transactions in cryptocurrency trading, and then summarize recent research on malicious transaction detection.

\subsection{Malicious Transaction Behavior}
In recent years, with the development of blockchain technology, a series of cryptocurrencies represented by Bitcoin and Ethereum have attracted significant attention. As of July 2024, the market value of cryptocurrencies has exceeded \$2.3 trillion, surpassing silver and becoming the eighth largest asset in the world. The significant wealth effect and the decentralized and anonymous characteristics of its technology make cryptocurrencies more susceptible to becoming a breeding ground for malicious activities. malicious activities under the pretext of digital economy and technological innovation are on the rise. The most common malicious activities currently include Phishing transactions, Money laundering, Ponzi schemes, Blackmail, Darknet market and others.

\textbf{Phishing scam} \cite{TIFS24Fishing}. It manifests as unauthorized transactions caused by stolen private keys or credentials. A key characteristic is the occurrence of unusual and unauthorized withdrawals, usually happening shortly after the credentials are stolen. These transactions typically occur with a low frequency but involve large amounts, with the compromised account sending funds to addresses controlled by the fraudster.

\textbf{Money laundering} \cite{TIFS24ML}. It manifests through the circulation of funds back to the original owner, with transactions happening in rapid succession and at a very high frequency. This activity usually involves a small group of colluding entities or a single entity using multiple accounts to create the illusion of market activity.

\textbf{Ponzi schemes} \cite{WWW18}. A fraudulent investing scam promising high returns with little risk, where returns are paid to earlier investors using new investors' funds. Ponzi schemes are characterized by an initial inflow of funds from new investors, which is used to pay returns to earlier investors, creating an illusion of profitability. Transactions typically show a high frequency of new investments followed by consistent withdrawals by earlier participants. 

\textbf{Blackmail} \cite{MFScope}. It has three typical categories that are ransomware, sextortion, and scam. The common characteristic of these behaviors is sending large one-time transactions to the attacker's address, usually occurring shortly after the extortion request is made. These transactions are typically one-time or very few in number, with the victim transferring large amounts of funds directly to the attacker.

\textbf{Darknet market} \cite{MFScope}. It is characterized by funds flowing from buyers to sellers through multiple intermediary addresses, often to obfuscate the origin of funds. These transactions occur irregularly, depending on market activity and availability of goods/services. The frequency varies widely, influenced by demand and law enforcement actions. The entities involved include anonymous buyers, sellers, and market administrators operating on darknet platforms.

Due to the rapid evolution of malicious transaction behaviors, in addition to the common malicious transactions mentioned above, there are also new types of malicious transactions such as Rug Pulls \cite{RugPull}, Flash Loan\cite{FlashLoan}. These malicious transactions fundamentally involve users transferring funds to related parties for profit through covert means. However, due to the lack of labeled data for these behaviors, they are not the focus of this paper.

\subsection{Related Work}
In this section, we systematically introduce recent research work on the detection of malicious transactions in cryptocurrencies. We categorize the existing research into four types based on different technical approaches: empirical analysis, machine learning, and graph learning, as shown in Table \ref{tab:related work}.

\textbf{Empirical Analysis.} This type of research mainly relies on expert experience, identifying malicious transactions and addresses from massive transaction data based on the characteristics and behavior patterns of malicious transactions. He et al. \cite{CCS23} detected phishing transactions by dynamically accessing suspicious websites and simulating phishing transaction outcomes. Liu et al. \cite{TIFS24Fishing}  further expanded the detection scope of phishing addresses by utilizing the interaction relationships between phishing addresses and achieved group detection of phishing gangs using community optimization algorithms.

\textbf{Machine Learning.} This type of method mainly extracts statistical features of malicious transactions from aspects such as transaction amount, time, and frequency, and trains classifiers using traditional machine learning methods to achieve malicious transaction detection. Chen et al. \cite{WWW18} and Zheng et al. \cite{MulCas} combined opcode features smart contracts with user account features, using XGBoost to construct a Ponzi scheme detection method. Hu et al. \cite{Bert4ETH} achieved pre-training of the classification model using unlabeled data based on the BERT framework, and then fine-tuned it with a small amount of phishing data to obtain the phishing detection model.

\textbf{Graph Learning.} This type of research mainly uses graph embedding algorithms to convert malicious transaction graphs into low-dimensional vector representations, combined with machine learning classification algorithms to achieve malicious transaction detection. Jin et al. \cite{JSAC22} utilized GraphSAGE to aggregate the topological features of the transaction graph into the central transaction node. In addition to considering the topological features, Wu et al. \cite{TSMC22} and Li et al. \cite{TTAGN} also considered the transaction time information within the graph. They designed biased random walk strategies and attention-weighted mechanisms, respectively, to aggregate the transaction time information of neighboring nodes to the central node.

\textbf{Summary.} The limitations of existing methods lie in three aspects. First, empirical analysis and machine learning methods rely on prior knowledge and handcrafted features based on expert experience. Although effective in identifying specific malicious behaviors, their limited generalization reduces their effectiveness when confronted with diverse and rapidly evolving malicious activities. Second, while graph-based learning methods significantly enhance the recognition of malicious behaviors by considering the interconnectedness of transactions, their reliance on specific network structures and transaction features results in a lack of robustness and portability across different datasets and blockchain platforms. Finally, existing methods often depend on labeled data, but the limited availability of such datasets in real-world scenarios significantly restricts their effectiveness.

\section{Problem Definition}\label{subsec:Designgoal}
This paper mainly focuses on identifying accounts in cryptocurrency via contrastive learning, specifically from the perspective of node classification. The transaction graph constructed from cryptocurrency transaction data is usually represented as $G=(V,E,X)$, where $V=\{v_{1},v_{2},\ldots,v_{n}\}$ is the set of account nodes, $E=\{v_{i},v_{j}| v_{i},v_{j} \in V\}$ is the set of interaction edges, and $X \in \mathbb{R}^{m \times n}$is the feature matrix to represent the information of the interaction features of each account, where $m$ and $n$ denote the number of accounts and features. Without loss of generality, we assume that there are \( k \) types of malicious accounts, $Y=\{y_{1},y_{2},\ldots,y_{k}\}$, each representing a distinct class of malicious behavior. Based on the above definitions, the design goals of the malicious transaction detection method are as follows:

\textbf{Classifiable across various malicious behaviors.} Given the labeled accounts dataset \( P_1 = \{v_1, \ldots, v_p\} \) with their respective classes \( y_1, \ldots, y_k \), the detection method aims to classify the remaining addresses \( U_1 = \{v_{p+1}, \ldots, v_m\} \). Specifically, it aims to predict \( Y_{v_{p+1}, \ldots, v_{m}} \subseteq \{y_{1},y_{2},\ldots,y_{k}\} \).

\textbf{Transferable to new malicious transactions.} Given the labeled accounts dataset \( P_2 = \{v_1, \ldots, v_p\} \) with their respective classes \( y_1, \ldots, y_k \), the detection method aims to classify the remaining addresses \( U_2 = \{v_{p+1}, \ldots, v_m\} \). Specifically, it aims to predict \( Y_{v_{p+1}, \ldots, v_{m}} \subseteq \{y_{1},y_{2},\ldots,y_{k+l}\} \), where $l$ is the number of types of malicious transactions that the method has never seen.

\textbf{Transferable to new platform.} Given transaction graphs $G_i = (V_i, E_i, X_i)$ from blockchain platform $P_i$ and $G_j = (V_j, E_j, X_j)$ from an unseen platform $P_j$, the detection method trained on $G_i$ aims to generalize to $G_j$. Specifically, it aims to accurately classify account $v \in V_j$ as malicious or not based on the learned features from $G_i$.

\begin{figure*}[htb]  
	\centering  
	\includegraphics[width=\textwidth]{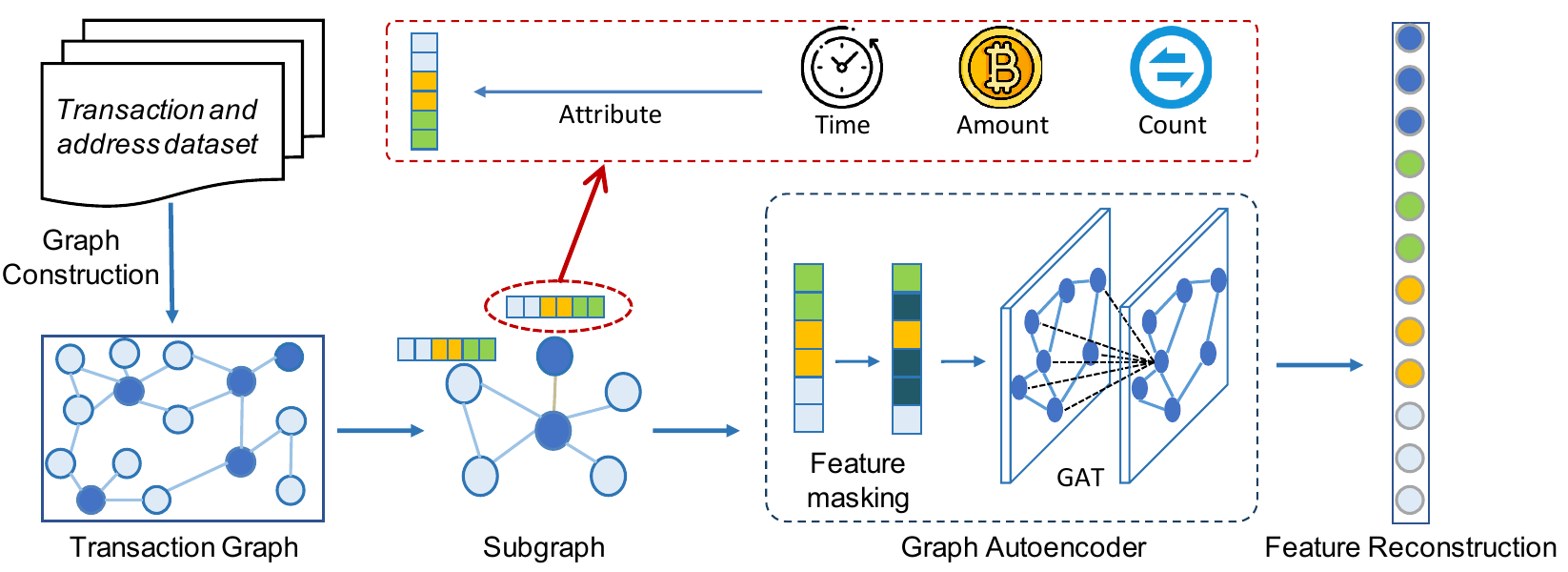}  
	\caption{Constructing TxGraph through transaction raw data.}  
	\label{fig: representation}  
	\vspace{-0.2cm}  
\end{figure*}

\section{Transaction Representation}\label{sec:representation}
In this section, we present transaction representation Transaction Graph (TxGraph) as the representation of malicious transactions and their topology information in transaction graph. Then, we describe the method for constructing TxGraphs from raw data. Finally, we use quantitative methods to verify the effectiveness of TxGraph. The construction process of TxGraph is depicted in Fig. \ref{fig: representation}.

\subsection{Transaction Attribute of Node}
Node attributes, which reflect an account's transaction habits in terms of time and amount, as well as its interactions with neighboring nodes, are a crucial component of TxGraph. These attributes primarily include transaction temporal, transaction amount, and transaction counts, as shown in Table \ref{tab:account_attribute}.

\begin{table}[htb]  
    \centering  
    \caption{Introduction to Account Attributes.}  
    \label{tab:account_attribute}  
    \renewcommand\arraystretch{1.2}  
    \begin{tabular}{@{}llp{0.5\linewidth}@{}}
        \toprule  
        Category & Number & Description \\  
        \midrule  
        Temporal & $f_1-f_{17}$ (17) & Temporal attributes such as account lifecycle, active periods, and transaction intervals \\ 
        \midrule  
        Amount & $f_{18}-f_{36}$ (19) & Amount attributes such as total output amount and the difference between output and input amounts \\  
        \midrule  
        Counts & $f_{37}-f_{43}$ (7) & The attributes of the target node in the transaction graph related to the number of transactions of its neighbors \\  
        \bottomrule  
    \end{tabular}  
\end{table}

1) Transaction temporal. Malicious transaction behaviors often exhibit certain temporal characteristics. For instance, addresses involved in market manipulation might engage in high-frequency trading to influence market prices \cite{MarketManipulation}, some phishing websites may be active only during specific holidays or special events \cite{CCS23}, and money laundering addresses may conduct transactions periodically \cite{TIFS24ML}. Therefore, we consider temporal attributes such as the active period of the address and the average transaction interval.

2) Transaction amount. Analysis of transaction amounts can also uncover malicious transaction behaviors. Money laundering often involves numerous small transactions to obscure the source and destination of funds \cite{TIFS24ML}, while phishing activities usually result in a few large transactions \cite{CCS23}. Therefore, We considered amount attributes such as the total input of the address and the difference between the maximum and minimum single transaction inputs.

3) Transaction counts. It is represented as node degree information in the transaction graph, reflecting the interaction between malicious accounts and their neighboring nodes. For example, phishing accounts generally receive funds from a few sources, resulting in relatively low in-degree. Ponzi scheme accounts receive funds from a large number of investors, resulting in a very high in-degree. Therefore, we consider degree-related attributes such as in-degree, out-degree, and the difference between them.

\begin{table}[htb]
	\begin{center}
		\caption{Notations used in this paper.}
		\label{tab:notations}
		\renewcommand\arraystretch{1.1}
		\resizebox{0.7\linewidth}{!}{
			\begin{tabular}{c r}
				\hline
				Notation & Description \\
				\hline
				$G$ & Transaction graph \\
				$v$ & Transaction Node \\
                $V$ & Node set\\
				$E$ & Edge set \\
				$x$ & Node feature\\
                $X$ & Node feature matrix\\
                $z$ & Graph structure feature\\
                $W$ & Weight parameter\\
                $N(v)$ & The neighbor nodes of v\\
                $h$ & Feature representation of node \\
                $\alpha$ & Coefficient of attention \\
                $y$ & Address label\\
                $\mathcal{L}$ & The loss function\\
                \hline
		\end{tabular}}
	\end{center}
 \vspace{-0.2cm}
\end{table}

\subsection{Transaction Graph}
We propose Transaction Graph (TxGraph) as a novel transaction representation. Specifically, we combine the transaction attributes of account nodes and use a graph autoencoder (GAE) to capture the topological information between account nodes. To ensure consistency in the contributions of different transaction attributes to structural feature computation, we normalize the transaction attributes. Table \ref{tab:notations} summarizes the main symbols used in this paper.

Exclusively relying on transaction attributes of nodes neglects the topological information regarding interactions between nodes and their neighbors, thus inadequately portraying malicious transactions. For instance, in the dark web, money laundering transactions frequently employ a series of intermediary addresses for multiple transfers to obscure the source and ultimate purpose of funds \cite{MFScope}. In such cases, relying solely on statistical features of each address (e.g., transaction amount, time, counts) can be misleading, as each address may superficially appear engaged in normal transactions. However, considering the topological structure, particularly the transaction relationships between addresses, is crucial for uncovering these chain-like transfer patterns indicative of malicious transactions.

To extract more robust structural features, we adopt the Graph Autoencoder (GAE) as the training paradigm for the graph feature extractor. GAE predicts masked node attributes using autoregressive methods, enabling the extractor to learn latent patterns and structural information from the data. This minimizes noise impact and enhances method robustness and generalization capability \cite{VGAE}. Furthermore, we select the Graph Attention Network (GAT) as the graph feature extractor. Compared to GCN and GraphSAGE, GAT employs attention mechanisms to dynamically learn weights based on node relationships, enabling precise capture of subtle differences between nodes \cite{GAT}. This makes GAT more suitable for our node classification task.

Specifically, GAT first calculates the attention coefficient \( \alpha_{vu} \) between node \( v \) and each of its neighbors \( u \), which determines the importance of the neighbor nodes when aggregating the features of the neighbor nodes. This parameter is calculated using the following formula:

\begin{equation} 
    \alpha'_{uv} = \frac{\exp(LeakyReLU(e'_{uv}))}{\sum_{k \in N_u} \exp(LeakyReLU(e'_{uk}))} 
\end{equation}
Where \(LeakyReLU\) is a rectified activation unit with a small negative slope, and the \(softmax\) function is used to normalize the attention coefficients. The calculation of \(e'_{uv}\) is as follows:

\begin{equation}
	e'_{uv} = a(\left[ W h_u \,||\, W h_v \right]), v \in N_u 
\end{equation}
Where the linear transformation parameter \( W \) is used for the linear transformation of the feature dimensions of the nodes, \(h_v\) and \(h_u\) are the feature vectors of nodes \(v\) and \(u\) respectively. \(\|\)denotes the concatenation operation, and \( a \) maps the concatenated high-dimensional features to a real number.

Subsequently, each node aggregates its neighbors' features weighted by the corresponding attention coefficients. This process is calculated as described by the following formula:

\begin{equation}
	h_v^{(k)} = \sigma\left( \sum_{u \in \mathcal{N}(v)} \alpha_{vu}^{(k)} W^{(k)} h_u^{(k-1)} \right)
\end{equation}
where \( h_v^{(k)} \) is the feature representation of node  \( v \) in the \( k \)-th layer, \( \mathcal{N}(v) \) is the set of neighbor nodes of \( v \), \( W^{(k)} \) is the weight matrix of the \( k \)-th layer, \( \alpha_{vu}^{(k)} \) is the attention coefficient, and \( \sigma \) is the nonlinear activation function.

Next, we reconstruct the adjacency matrix $A$ from the latent embeddings $Z$ by inner product decoder:

\begin{equation}
    \hat{A} = \sigma (ZZ^\top)
\end{equation}
where $\hat{A}$ is the reconstructed adjacency matrix obtained after decoding, and $\sigma$ is the sigmoid function applied element-wise.

To reconstruct the node features, we use the mean squared error (MSE) loss function, which measures the difference between the original node feature matrix \(X\) and the reconstructed node feature matrix \(\hat{X}\):

\begin{equation}
\mathcal{L}_g = \frac{1}{N} \sum_{i=1}^{N} \|\mathbf{x}_i - \hat{\mathbf{x}}_i\|^2
\end{equation}
where $N$ is the number of transaction nodes, \(\mathbf{x}_i\) is the original feature vector of node \(i\) and \(\hat{\mathbf{x}}_i\) is the reconstructed feature vector of node \(i\).

Lastly, we combine node transaction attributes with topological features to achieve a comprehensive representation of node transactions:

\begin{equation}
\tilde{x}_i=\mathrm{CONCAT}(x_i, h_v^{(k)})
\end{equation}
where $\tilde{x}_i$ represents the node features after feature fusion.

\subsection{Evaluation on Effectiveness of TxGraph}
We use quantitative metrics to assess the effectiveness of TxGraph and the two typical representations used by SOTA methods: statistical feature-based representation  (note as ST.) \cite{ESWA20} and graph feature-based representation (note as Gra.) \cite{TNSE24}.

We chose six representative malicious transaction from the publicly available dataset BABD-13 \cite{2024TIFS-Dataset}. For each type of malicious transaction, we randomly select 15 transaction samples and transform them into corresponding transaction representations. Then we use the euclidean distance to assess the dissimilarity between different malicious transactions. A greater distance indicates a more effective capture of the distinguishing features between malicious transactions. To allow data to be scaled to fall within defined intervals without changing the original data source distribution and comparable with each other, we take the average of all distances and use min-max normalization.

\begin{figure}[htbp]
	\small
	\centering
	\includegraphics[width=\linewidth]{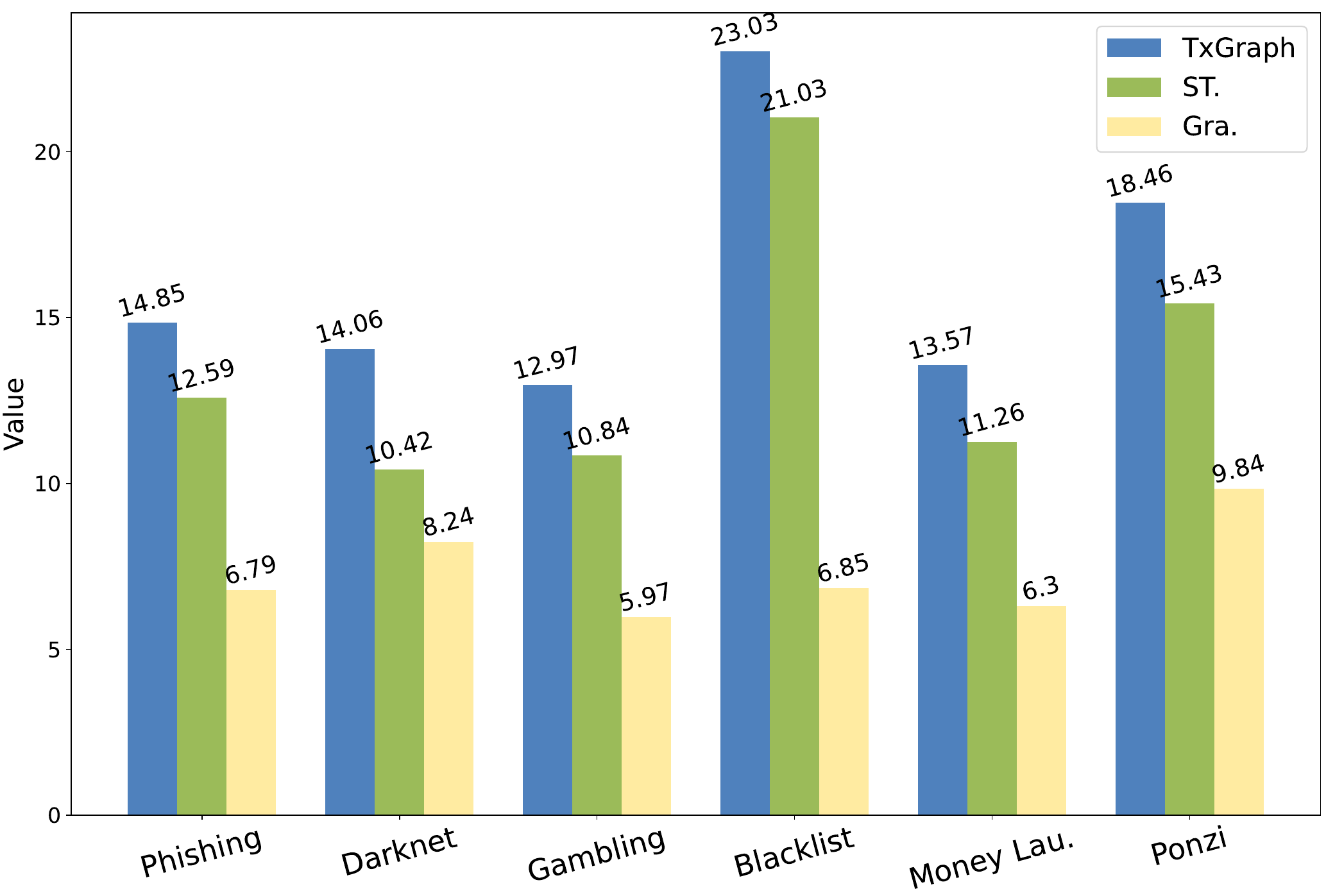}
	\caption{The distance between different malicious transactions under different representations.}
	\label{fig:Reprsentation-Tx}
\end{figure}

As shown in Fig. \ref{fig:Reprsentation-Tx}, TxGraph exhibits the maximum distance on average in the six categories of malicious transactions, demonstrating its ability to distinguish between different malicious transactions. Gra. representation has the smallest distance on average, indicating that the graph feature-based representation is less expressive.

\textbf{Evaluation on the across-platform scenarios.} In real-world scenarios, malicious actors \cite{Usenix19} might exploit the structural and operational differences between cryptocurrency platforms to evade detection. We investigate the importance of across-platform detection by focusing on dominant cryptocurrency platforms: Bitcoin \cite{2024TIFS-Dataset} and Ethereum \cite{TIFS24Fishing,TIFS24ML}. The distinct characteristics of these platforms present unique challenges for malicious detection. For instance, criminal might leverage Bitcoin's simpler transaction method to perform rapid fund transfers \cite{MFScope}, while utilizing Ethereum's complex smart contract interactions to obscure malicious activities \cite{JSAC22}. 

\begin{figure}[htbp]
	\small
	\centering
	\includegraphics[width=\linewidth]{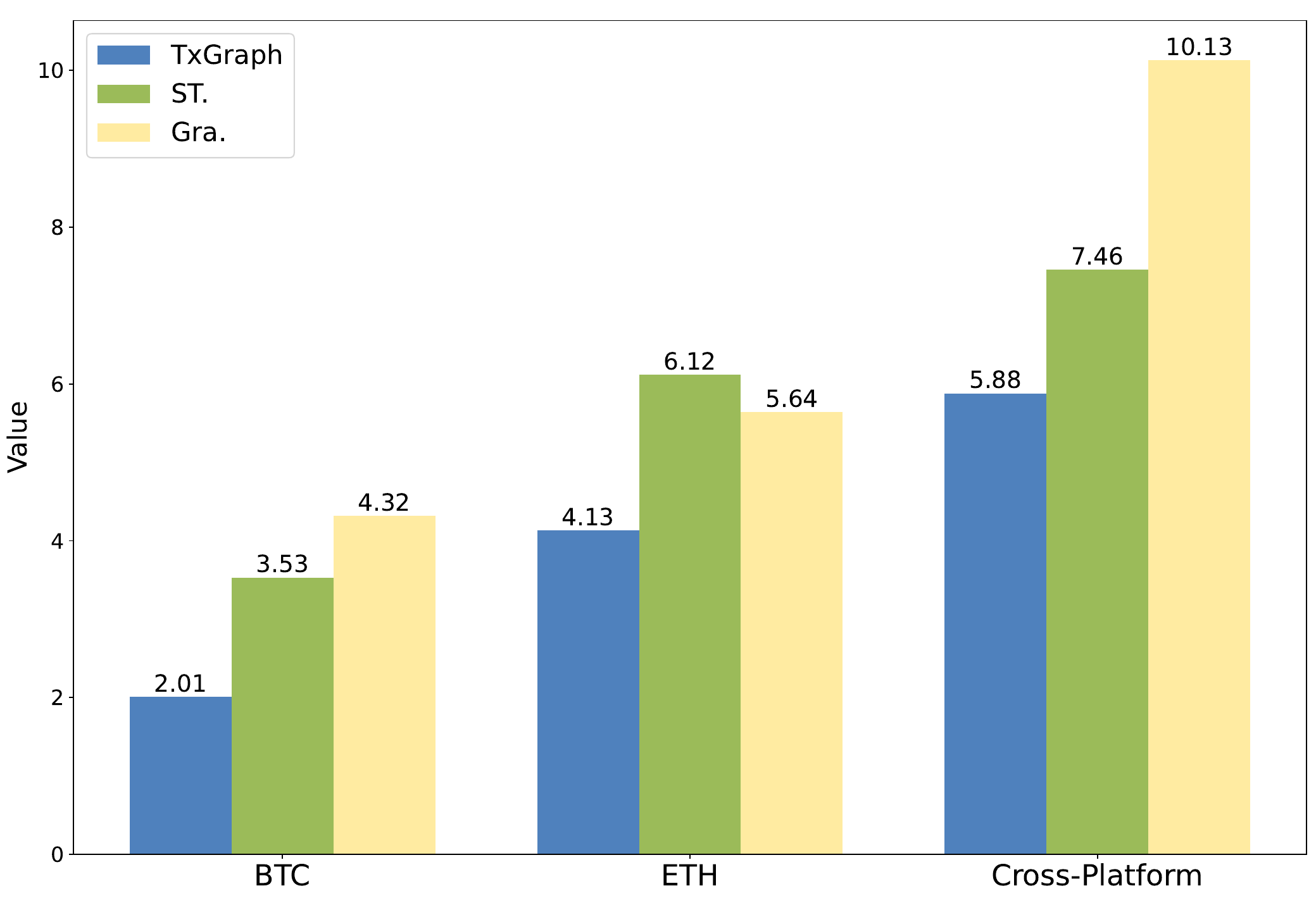}
	\caption{The distance between different malicious transactions on BTC and ETH under different representations.}
	\label{fig: representation-platform}
\end{figure}

Besides Bitcoin's malicious transaction data, we selecte 15 malicious transaction samples from the Ethereum dataset, which will be introduced in Section \ref{sec:Experiments}. We then calculate the distances between malicious transaction representations within each platform and between the two platforms. As shown in Figure \ref{fig: representation-platform}, TxGraph consistently exhibits the smallest distance compared to the other two representation methods in all cases. This is because TxGraph captures robust transaction features for both BTC and ETH, demonstrating its effectiveness in identifying across-platform malicious transactions.

\section{The Proposed ShadowEyes}\label{sec:ShadowEyes}
\begin{figure*}[htb]
	\centering
	\includegraphics[width=\textwidth]{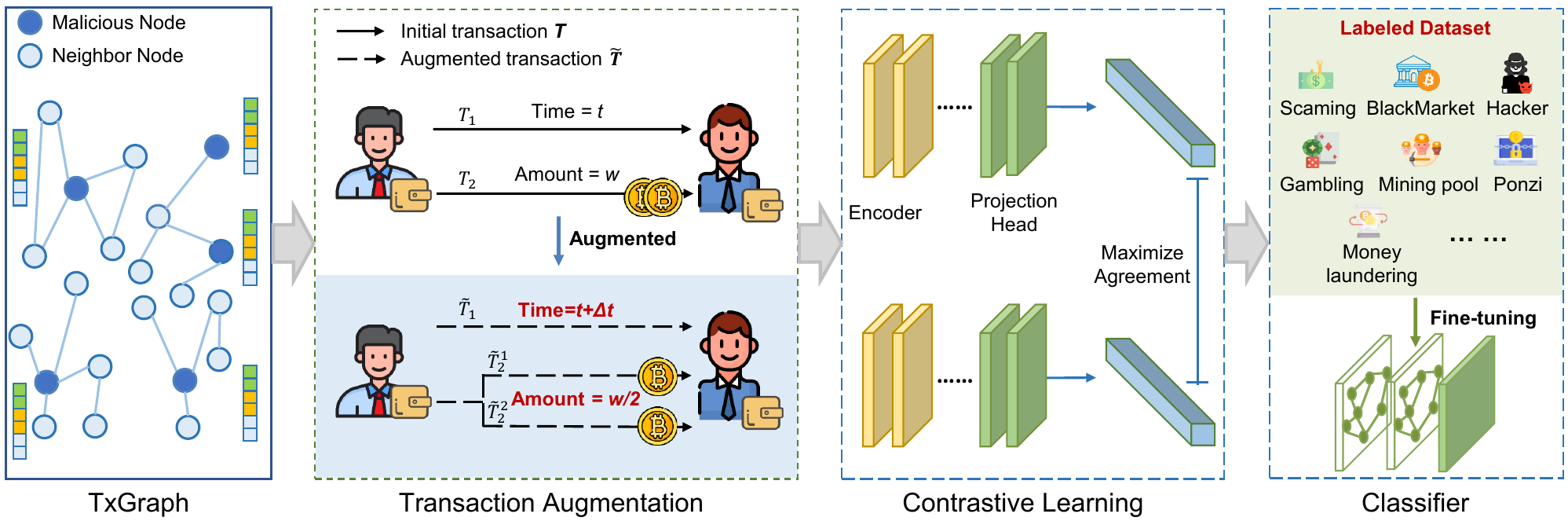}
	\caption{The system overview of ShadowEyes.}
	\label{fig:ShadowEyes}
	\vspace{-0.2cm}
\end{figure*}

With the help of Txgraph, the detection of malicious transaction s is turned into a graph classification problem. In this section, we present the design details of ShadowEyes, a method for detecting malicious transactions based on graph representation and contrastive learning. The overview of the ShadowEyes is shown in Fig. \ref{fig:ShadowEyes}.

\subsection{Method Overview}\label{subsec:Model}
The primary components of ShadowEyes consist of  the \emph{Transaction Representation} module, the \emph{Transaction Augmentation} module, the \emph{Contrastive Learning} module, and the \emph{Classifier} module.

\textbf{Transaction Representation}. As outlined in \ref{sec:representation}, this module generates a generalized graph structure representation, called TxGraph, by extracting interaction features between the target account and its neighboring accounts. TxGraph is then passed to subsequent modules for data augmentation.

\textbf{Transaction Augmentation}. This module simulates the operations of malicious transactions to evade regulation by generating augmented samples through adding random time delays and splitting transaction amounts. Both the original and enhanced transaction samples are used as inputs for the subsequent contrastive learning module.

\textbf{Contrastive Learning}. This module leverages unlabeled transaction data to train an encoder to map transaction samples into an embedding space. The trained encoder can ensure the similarity between original transactions and their corresponding augmented transactions.

\textbf{Classifier}. This module fine-tunes a classifier using the pre-trained encoder and a small amount of labeled samples. For malicious transaction detection, test samples are fed into the module to obtain detection results.

\subsection{Transaction Augmentation}\label{subsec:data augmentation}
Graph augmentation enhances the diversity of training data by applying random transformations and perturbations to existing samples without altering their labels, thereby improving a method's ability to generalize. In computer vision, typical data augmentation techniques include rotating, cropping, color transformations, and flipping images \cite{CVPR2019}. However, these methods are less effective for malicious transaction detection. Since account behaviors are represented by vectors containing semantic information, applying transformations like rotations, croppings, and flips to these vectors does not yield meaningful results and fails to increase the diversity of the sample data.

In malicious transaction detecting scenarios, effective data augmentation methods should follow two rules: (1) \emph{Label invariance}. The behavior samples of the augmented address may deviate to some extent in the feature space, but their address behavior remains unchanged. (2) \emph{Randomness and diversity}. The augmentation methods should maximize data diversity to simulate various variants of malicious transaction behaviors. Based on this, we propose a novel data augmentation method to maintain the semantics of the augmented samples.

\textbf{Time random delay.} Given a user address $U$, a random delay of time length $ \Delta t $ is added to the timestamps of all their historical transactions with a certain probability $\mathcal{P}$, and $\mathcal{P}$ follows an i.i.d. uniform distribution. 

\begin{equation}
\hat{t}=t+ \Delta t 
\end{equation}
where $\hat{t}$ denote the modified timestamp for a transaction, and $t$ represents the initial timestamp.

In real-world scenarios, a user's transaction can experience random fluctuations in transaction times due to network transmission delays. During congested periods in blockchain networks, transactions may be delayed because of limited mining computational power. Therefore, applying data augmentation through random time delays can help the method learn more robust temporal features and mitigate overfitting. This approach adjusts the original user samples without altering the inherent category of the user.

\textbf{Transaction amount split.} Given a user address $U$, with a certain probability $\mathcal{P}$, select $ \theta $ transactions from its historical transactions. For each selected transaction $T$, split its amount equally into two new transactions ${\widetilde{T}}^1$ and ${\widetilde{T}}^2$ , adding a random time delay to ${\widetilde{T}}^2$:
\begin{equation}
t_2=t_1+ \Delta t 
\end{equation}
where $t_1$ represents the time of transaction ${\widetilde{T}}^1$, and $t_2$ represents the time of transaction ${\widetilde{T}}^2$.

\begin{figure}[htbp]
	\small
	\centering
	\includegraphics[width=\linewidth]{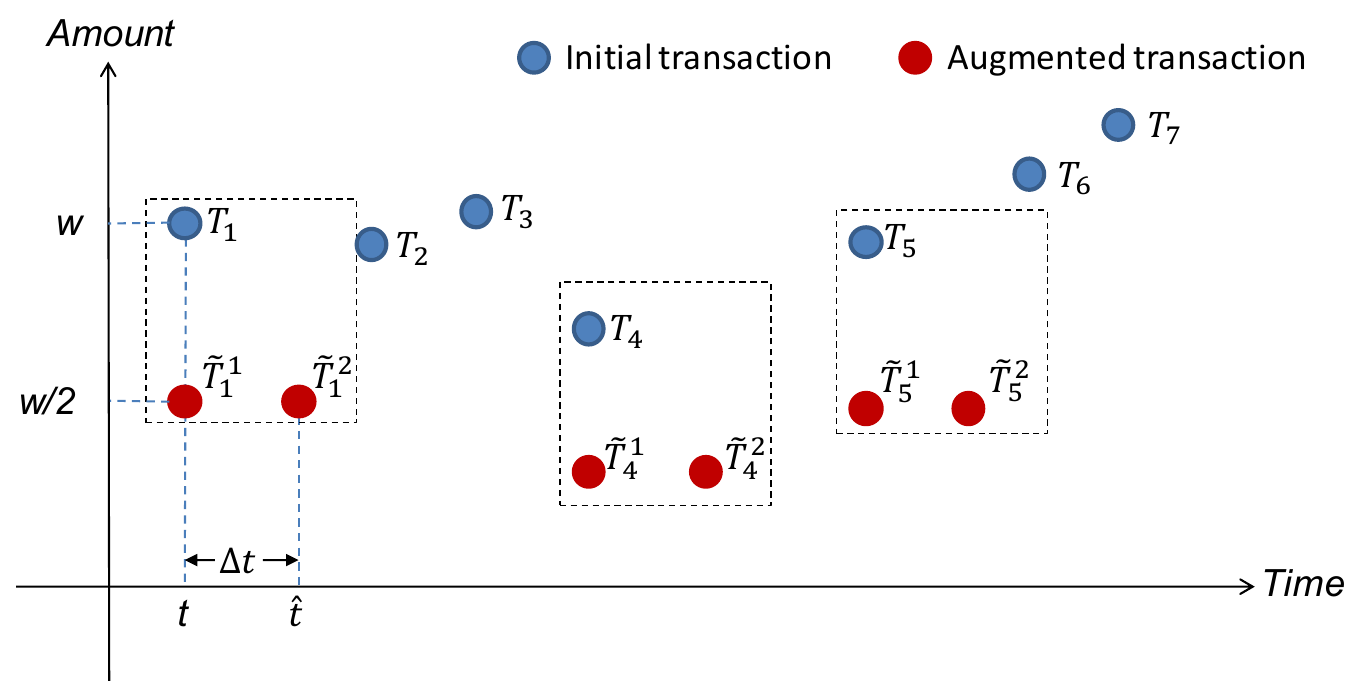}
	\caption{Transaction amount split.}
	\label{fig AU-amount}
        \vspace{-0.2cm}
\end{figure}

In real-world scenarios, malicious users often transfer large illegal assets through small transactions to avoid detection by regulators. By incorporating this behavior into the training data, the method can better identify such evasion strategies. As illustrated in Fig. \ref{fig AU-amount}, this data augmentation method simulates the behavior of users splitting large transactions into multiple smaller transactions. In the figure, blue circles indicate the original transactions, while red circles represent the augmented transactions resulting from the splitting operation.

After the augmentation opration, we generate two correlated views ${\widetilde{h}}^1$, ${\widetilde{h}}^2$ for given node reprentation $h$. These views serve as positive sample pairs and are used as pre-training samples for the encoder.

\subsection{Contrastive Learning}\label{subsec:Classifier}

After the data augmentation operation, the original data and the augmented samples are fed into a contrastive learning-based pre-training model. This model consists of an encoder and a projection head. The encoder learns the deep representations of user samples as inputs to the projection head, which performs dimensional transformation to output the final encoding. Guided by the contrastive loss, the model will try to make the encoding outputs of similar sample pairs closer and the encoding outputs of different samples further apart.

\begin{figure}[htbp]
	\small
	\centering
	\includegraphics[width=\linewidth]{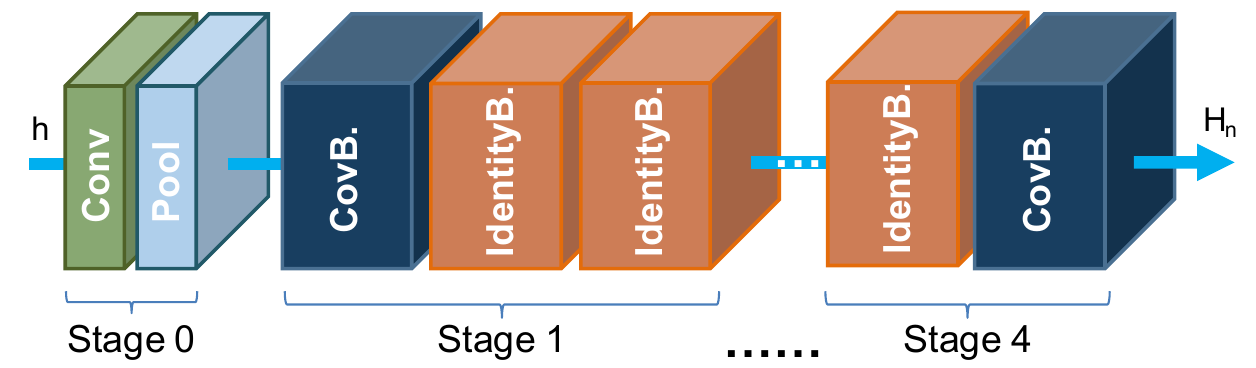}
	\caption{Structure of the encoder.}
	\label{fig: resnet-50}
        \vspace{-0.2cm}
\end{figure}

\begin{figure}[htbp]
    \centering
    \subfigure[Identity block]{
        \includegraphics[width=0.44\linewidth]{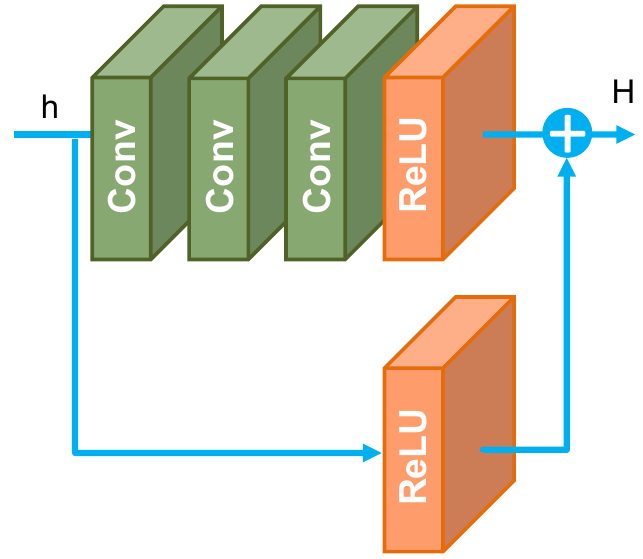}
    }
    \hfill
    \subfigure[Convolutional block]{
        \includegraphics[width=0.44\linewidth]{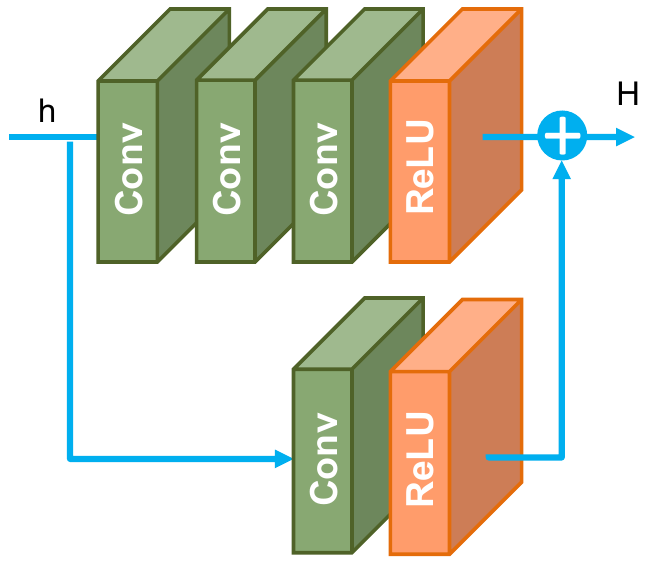}
    }
    \caption{Residual block in Resnet-50.}
    \label{fig:resnet-block}
\end{figure}

In this paper, we use ResNet-50 \cite{Resnet} as the foundational architecture for the encoder, as illustrated in Fig. \ref{fig: resnet-50}. ResNet-50 effectively addresses the issues of gradient vanishing and explosion that are common in deep networks by incorporating residual blocks. These blocks are designed to learn the residual mapping, which is the difference between the output and the input, rather than solely focusing on the output mapping of the preceding layer. In ResNet-50, there are two main types of residual blocks: the Convolutional Block and the Identity Block, as shown in Fig. \ref{fig:resnet-block}. The key difference between them is whether the skip connection includes an additional convolutional layer, which is used to align the output dimension of the residual block with the input dimension. The computation of a residual block for an input $h$ is represented as follows:

\begin{equation}
 H = {F}(h, \{W_a\}) + W_b {h}
\end{equation}
where ${F}(h, \{W_a\})$ denotes the transformation function within the residual block. $W_a$, $W_b$ are convolution kernels, and $W_b$ is used for dimensionality reduction or expansion of the input. Based on the residual block, the computation of ResNet-50 can be represented as:

\begin{equation}
	H_n = {h}_n + \sum_{i=1}^{n} {F}(h_{i-1}, \{W_i\})
\end{equation}
where $H_{n}$ is the output of the $n$-th residual block, and ${F}({h}_{i-1}, \{W_i\})$ denotes the convolution operation of the $i$-th residual block. $n$ and $h_{n}$ denote the number of residual blocks and the input of the $n$-th residual block.

Then, a projection head module consisting of a bias-free fully connected layer, normalization layer, activation layer, and fully connected layer will project $y_{n}$ into the embedding space, yielding $s$ and $s'$. Finally, the goal of contrast is to maximize the consistency between positive pairs in the embedding space via minimizing the contrastive loss:

\begin{equation}
	\text{sim}(s,s') = \frac{s^T s'}{\|s\| \|s'\|}
\end{equation}

\begin{equation}
	\mathcal{L}_c = -\log \frac{\exp(\text{sim}(s_i, s_i'))}{\sum_{k=1}^N \exp(\text{sim}(s_i, s_k))}
\end{equation}
where $N$ denotes the number of samples in a batch, and $sim(s_i, s_i')$ is the cosine similarity between the view embedding $s_i$ and $s_i'$.

The downstream task after pre-training is to train a classification model. We fine-tune the model using a labeled addresses dataset. Specifically, we freeze the encoder parameters that were trained during pre-training and then add new fully connected layers to train the classifier. 




\section{Performance Evaluation}\label{sec:Experiments}
In this section, we conduct extensive experiments to evaluate the effectiveness of ShadowEyes. We first introduce the experiment and parameter settings. Then, we thoroughly compare ShadowEyes with state-of-the-art methods in four typical scenarios: zero-shot learning, across-platform detection, imbalanced dataset, and few-shot learning. Furthermore, we perform ablation experiments to assess the contributions of different submodules within ShadowEyes.

\subsection{Environmental Settings}\label{subsec:data}
\textbf{Environments.} All the experiments are conducted on a server with a 64-bit Windows operating system, configured with a 4-core Intel(R) Xeon(R) Silver 4110 CPU at 2.10GHz, 16GB RAM, 11GB VRAM, and a 2080Ti GPU. The software environment installed on this system includes Python 3.7 and Pytorch 1.13.1, with a 100GB SSD and Cuda version 11.7.

\begin{table}[htb]  
    \centering  
    \caption{Bitcoin Transaction Dataset.}  
    \label{tab:btc_data}  
    \renewcommand{\arraystretch}{1.1}  
    \begin{tabular}{l r}  
        \toprule  
        Category & Number \\  
        \midrule  
        Personal Wallet & 1,502 \\  
        Mining Pool & 1,580 \\  
        Network Service & 91,617 \\  
        Digital Financial Service & 9,489 \\  
        \midrule  
        Phishing & 8,686 \\  
        Gambling & 105,257 \\  
        Ponzi Scheme & 15 \\  
        Money Laundering & 16 \\  
        Criminal Blacklist & 27 \\  
        Darknet Transaction & 13,861 \\  
        \midrule  
        \textbf{Total Number} & 232,050 \\  
        \bottomrule  
    \end{tabular}  
\end{table}

\textbf{Datasets.} The datasets used in this paper include Bitcoin transaction dataset, Bitcoin address behavior dataset and Ethereum account behavior dataset. 
\begin{itemize}
\item Bitcoin transaction dataset. This paper uses the data acquisition method provided in the literature \cite{KDD23}, collecting block data through public API\footnote{https://chain.api.btc.com/v3/block} and further parsing it. We obtained transaction data from block heights 585,000 to 650,000 (from July 12, 2019, to May 26, 2021), including a total of 516,167,131 addresses and 713,703,239 transactions. These data will be further selected to construct transaction subgraphs for labeled addresses.

\item Bitcoin labeled address dataset. This dataset is derived from BABD-13 \cite{2024TIFS-Dataset}. To eliminate the impact of over-representation of certain address behavior types, which can lead to artificially high recognition accuracy during the training phase, we removed 300,000 centralized exchange address entries from the original dataset. We further merged the behavior labels of P2P Financial Infrastructure Service and P2P Financial Service into Digital Financial Services, aiming to prevent ShadowEyes from over-focusing on subtle behavioral differences. Table \ref{tab:btc_data} presents the details of this dataset.

\item Ethereum labeled account dataset. This dataset comes from a public dataset \cite{ETH-Dataset}, which includes 9,841 addresses, of which 2,179 are malicious scam addresses, and the remaining are normal addresses. Each address contains 50-dimensional features. This dataset was used for the evaluation of ShadowEyes's scalability.
\end{itemize}

\textbf{Methods in comparison.} In order to comprehensively assess the performance of ShadowEyes, we employ eight typical methods for comparison, which are briefly described as follows. To ensure a fair comparison, all methods have been fine-tuned to achieve optimal performance on the dataset used for evaluation.  

\begin{itemize}
\item Feature Inversion Contrastive Learning (FICL) \cite{ProCL}, which pretrains the encoder using a data augmentation approach that flips the user feature vectors, and then trains the classifier based on labeled data.
\item BERT \cite{Bert4ETH}, an autoregressive model grounded in a bidirectional Transformer architecture, facilitates model training by predicting the masked portions of the feature vectors.
\item Decision Tree (DT) \cite{DesicionTree}, which recursively selects features that have significant impact on identifying malicious transactions to partition the transaction dataset, until the stopping criteria are met. 
 \item K-Nearest Neighbor (KNN) \cite{Arxiv-KNN}, which predicts the classification of a new data point by considering the labels of the K nearest neighbor nodes to the target sample, assigning the majority class of these neighbors as the category of the new data point.
 \item Logistic Regression (LR) \cite{KDD23}, which maps the transaction features onto probabilities and ascertains the transaction category by comparing these derived probabilities against a predefined threshold.
 \item MultiLayer Perceptron(MLP) \cite{MLP}, which employs multiple fully connected layers for non-linear mapping, learning intricate patterns from transaction data features. 
 \item ABGRL \cite{TNSE24}, which constructs multiple decision trees, trains them using random feature subsets, and combines their results through a voting mechanism to determine the final classification.
 \item CNN \cite{CNN}, which extracts local features from transaction data through hierarchical convolutional operations, enhances features and reduces dimensionality using pooling layers, and finally classifies transactions using fully connected layers.
 \end{itemize}

\textbf{Metrics.} We adopt three standard metrics for assessment, namely Precision, Recall, and F1-Score. Precision is computed as $\left( TP \right) / \left( TP + FP \right) $ and Recall is computed as $\left( TP \right) / \left( TP + FN \right) $. A high Precision indicates a reduced likelihood of misclassifying negative samples as positive ones, while a high Recall indicates a better ability of the method to identify truly positive samples. The F1-Score serves as a harmonic average of the method's precision and recall, defined as follows:

\begin{equation}
	\label{F1}
	\textrm{F1-Score}=\frac{2 \cdot \textrm{Precision} \cdot \textrm{Recall}}{\textrm{Precision} + \textrm{Recall}}
\end{equation}

\begin{table*}[ht]  
	\tiny  
	\setlength{\abovecaptionskip}{0cm}  
	\setlength{\belowcaptionskip}{-0.2cm}  
	\caption{Zero-Shot Learning Comparison of Different Methods.}  
	\label{tab:Zero-Shot}  
	\renewcommand\arraystretch{1.1}  
        \begin{threeparttable}
	\resizebox{\textwidth}{!}{  
		\begin{tabular}{clccccccccccc}  
			\hline  
			\multirow{2}{*}{Masked behavior} & \multirow{2}{*}{Metric} & \multicolumn{9}{c}{Method} \\ \cline{3-11}  
			& & CNN & ABGRL & DT & KNN & LR & MLP & FICL & BERT & \textbf{ShadowEyes} \\  
			\hline  
			\multirow{3}{*}{Phishing}  
			& Pre($\%$) & 80.68 & 89.74 & 81.30 & 75.49 & 77.42 & 77.48 & 70.83 & 86.28 & \textbf{91.19} \textcolor{red}{\scalebox{0.7}{$\blacktriangle$ 1.91}} \\  
			& Recall($\%$) & 75.18 & 75.72 & 70.92 & 69.09 & 59.92 & 63.07 & 66.93 & 83.58 & \textbf{84.79} \textcolor{red}{\scalebox{0.7}{$\blacktriangle$ 1.21}} \\  
			& F1($\%$) & 77.83 & 82.14 & 75.76 & 72.15 & 67.56 & 69.54 & 68.82 & 84.91 & \textbf{87.87} \textcolor{red}{\scalebox{0.7}{$\blacktriangle$ 2.96}} \\  
			\hline  
			\multirow{3}{*}{Darknet Market}  
			& Pre($\%$) & 75.28 & 85.69 & 76.96 & 71.21 & 72.50 & 73.57 & 71.22 & 82.04 & \textbf{89.57} \textcolor{red}{\scalebox{0.7}{$\blacktriangle$ 7.53}} \\  
			& Recall($\%$) & 68.55 & 80.12 & 62.90 & 64.26 & 52.93 & 58.22 & 68.01 & 78.37 & \textbf{85.88} \textcolor{red}{\scalebox{0.7}{$\blacktriangle$ 7.51}} \\  
			& F1($\%$) & 71.76 & 82.81 & 69.22 & 67.56 & 61.19 & 65.00 & 69.58 & 80.16 & \textbf{87.69} \textcolor{red}{\scalebox{0.7}{$\blacktriangle$ 7.53}} \\  
			\hline  
			\multirow{3}{*}{Gambling}  
			& Pre($\%$) & 63.82 & 72.08 & 55.07 & 60.51 & 58.70 & 60.49 & 54.30 & 71.65 & \textbf{82.57} \textcolor{red}{\scalebox{0.7}{$\blacktriangle$ 10.92}} \\  
			& Recall($\%$) & 54.11 & 68.86 & 53.18 & 45.92 & 41.68 & 42.26 & 46.01 & 59.36 & \textbf{72.11} \textcolor{red}{\scalebox{0.7}{$\blacktriangle$ 12.75}} \\  
			& F1($\%$) & 58.57 & 70.43 & 54.11 & 52.21 & 48.75 & 49.76 & 49.81 & 64.93 & \textbf{76.98} \textcolor{red}{\scalebox{0.7}{$\blacktriangle$ 12.05}} \\  
			\hline  
			\multirow{3}{*}{Blacklist}  
			& Pre($\%$) & 78.83 & 81.20 & 81.35 & 76.46 & 77.74 & 78.26 & 65.77 & 84.41 & \textbf{90.30} \textcolor{red}{\scalebox{0.7}{$\blacktriangle$ 5.89}} \\  
			& Recall($\%$) & 70.62 & 75.72 & 69.32 & 69.40 & 58.76 & 60.75 & 65.52 & 78.40 & \textbf{79.55} \textcolor{red}{\scalebox{0.7}{$\blacktriangle$ 1.15}} \\  
			& F1($\%$) & 74.50 & 78.36 & 74.85 & 72.76 & 66.93 & 68.40 & 65.64 & 81.29 & \textbf{84.58} \textcolor{red}{\scalebox{0.7}{$\blacktriangle$ 3.29}} \\  
			\hline  
			\multirow{3}{*}{Money Laundering}  
			& Pre($\%$) & 79.95 & 81.48 & 81.71 & 76.52 & 78.01 & 78.75 & 71.25 & 83.25 & \textbf{89.36} \textcolor{red}{\scalebox{0.7}{$\blacktriangle$ 6.11}} \\  
			& Recall($\%$) & 70.56 & 75.13 & 72.14 & 72.17 & 61.44 & 60.97 & 70.09 & 78.23 & \textbf{79.17} \textcolor{red}{\scalebox{0.7}{$\blacktriangle$ 0.94}} \\  
			& F1($\%$) & 74.96 & 78.17 & 76.63 & 74.28 & 68.74 & 68.73 & 70.67 & 80.66 & \textbf{83.96} \textcolor{red}{\scalebox{0.7}{$\blacktriangle$ 3.30}} \\  
			\hline  
			\multirow{3}{*}{Ponzi Scheme}  
			& Pre($\%$) & 79.17 & 81.09 & 81.59 & 76.31 & 78.09 & 78.52 & 73.96 & 83.73 & \textbf{90.25} \textcolor{red}{\scalebox{0.7}{$\blacktriangle$ 6.52}} \\  
			& Recall($\%$) & 71.00 & 76.85 & 70.45 & 74.17 & 61.60 & 64.70 & 73.38 & 77.01 & \textbf{77.95} \textcolor{red}{\scalebox{0.7}{$\blacktriangle$ 0.94}} \\  
			& F1($\%$) & 74.86 & 78.9 &75.61 &75.22 &68.87 &70.94 &73.67 &80.23 &\textbf{83.65}\textcolor{red}{\scalebox{0.7}{$\blacktriangle$ 3.42}}\\  
			\hline  
		\end{tabular}
	}
        \begin{tablenotes}
        \scriptsize       
        \item[1] We use \textcolor{red}{\scalebox{0.7}{$\blacktriangle$}} to highlight the improvement of classification performance compared with the SOTA BERT.
        \end{tablenotes}
        \end{threeparttable}
        \vspace{-0.2cm}
\end{table*}

\subsection{Evaluation on Zero-shot Learning}\label{subsec:Zero-Shot}
To evaluate the zero-shot learning capability of ShadowEyes, during the training phase, we randomly exclude one of the malicious types (Phishing, Darknet market, Gambling, Blacklist, Money laundering, and Ponzi scheme) from the original dataset. During the testing phase, the excluded malicious type is added back into the test dataset. We have the following observations through the results shown in Table \ref{tab:Zero-Shot}. 

1) ShadowEyes achieves the best performance across all metrics and excels in identifying various unknown malicious behaviors. Notably, in terms of F1-score, ShadowEyes shows a significant improvement, with the highest increase approaching 12\%. This indicates that ShadowEyes effectively captures complex patterns and relationships in transaction data through an unsupervised pre-training process, significantly enhancing the model's generalization ability.

2) BERT demonstrates suboptimal performance in most scenarios, indicating that while it can identify more potential malicious behaviors, it also generates a higher number of false positives. On the other hand, ShadowEyes effectively balances precision and recall, leading to higher F1 scores.

3) Methods such as KNN, ABGRL, and MLP can identify learned behavior types but are ineffective at recognizing new malicious transaction behaviors. As shown in the table, the accuracy of these methods is generally higher than their recall rates. This is because the effectiveness of supervised learning methods like KNN depends on the labeled data in the training set, resulting in limited cross-generalization.

4) When Gambling addresses are defined as unknown behaviors, the F1 scores of various methods generally become low outliers. This is because the feature distribution of gambling addresses is relatively concentrated in the training data, and the number of samples is higher compared to other malicious types. In this case, the models can more accurately learn the features related to gambling in the original training set, and after removing this type of data, the performance of each model generally fluctuates significantly.

\begin{figure}[htbp]
	\small
	\centering
	\includegraphics[width=\linewidth]{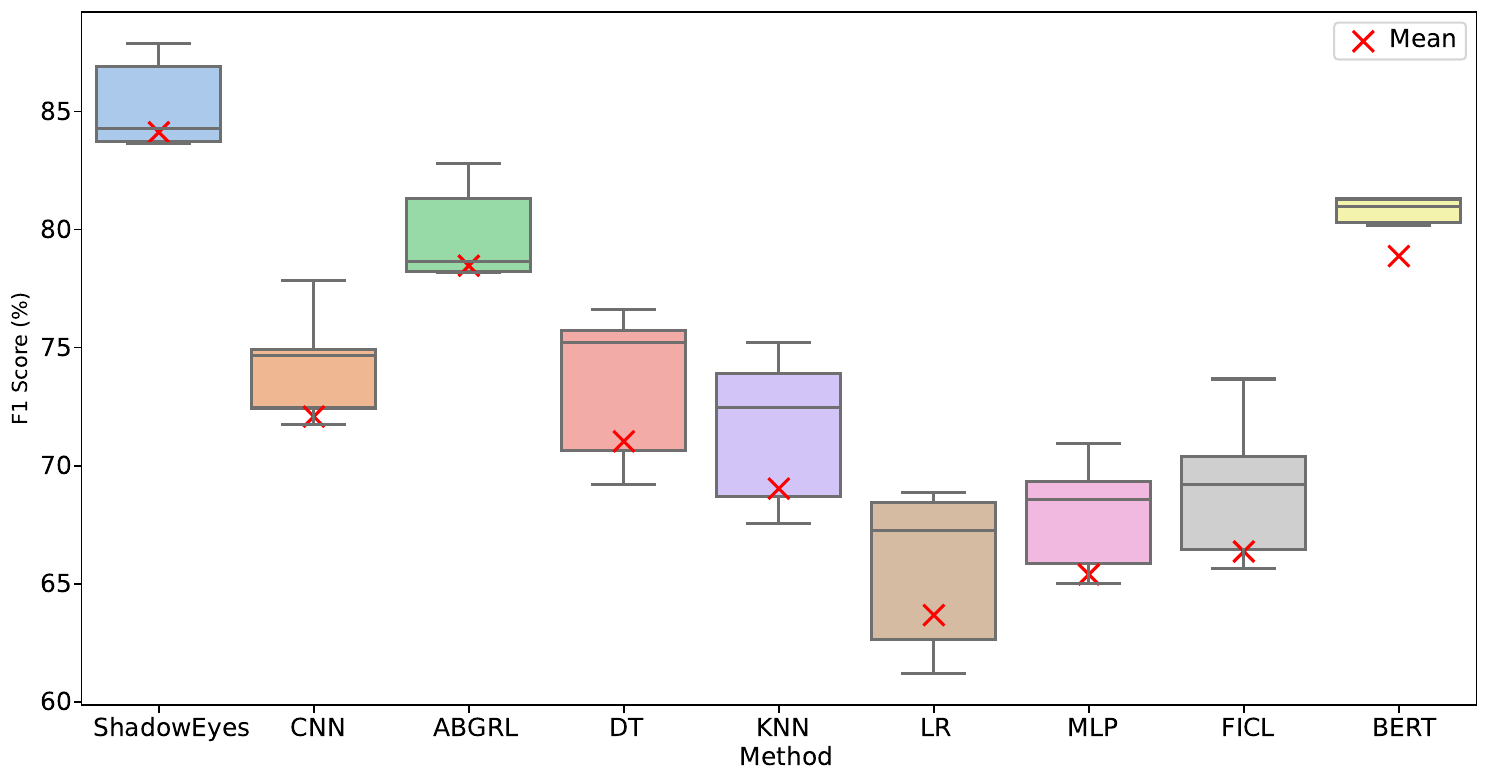}
	\caption{Distribution of F1 scores for each method in identifying unknown malicious behaviors.}
	\label{fig:box}
        \vspace{-0.2cm}
\end{figure}

We conducted an additional analysis to assess the model's performance stability in identifying diverse unknown malicious behaviors, as illustrated in Figure \ref{fig:box}. Each box plot represents the distribution of F1 scores for each method across different unknown behaviors, encompassing the maximum, 75th percentile, median (50th percentile), 25th percentile, and minimum values. The red cross denotes the average F1 score. From this figure, we can draw the following observations:

1) ShadowEyes exhibits the most stable performance in identifying various unseen malicious transactions. Among all methods, ShadowEyes has the highest median F1 score, close to 85\%, and its box plot range is relatively narrow, indicating stable and excellent performance in identifying unknown malicious behaviors.

2) The performance of pre-training methods depends on the quality of the training dataset in the pre-training stage. BERT's median is about 78\% with a relatively narrow box plot range, performing overall better than other traditional machine learning methods, indicating good generalization ability. The poor performance of FICL can be attributed to the fact that its data augmentation method does not effectively simulate actual malicious transaction behavior.

3) Machine learning methods do not require pre-training generally exhibit poor zero-shot learning ability. The box plots of various methods in the figure show a wide range, with average values typically around 75\%. This is because these methods rely heavily on labeled data for training, leading to limited robustness.

\subsection{Evaluation on Platform Scalability}\label{subsec:Scalability}
We evaluate the performance of the ShadowEyes method in address behavior classification across platforms. In the scenario of Bitcoin migrating to the Ethereum platform, ShadowEyes and the other two methods requiring pre-training are first pre-trained on the Bitcoin dataset, and then fine-tuned in the second stage using the Ethereum labeled account dataset. Other methods that do not require pre-training are directly trained using the Ethereum labeled account dataset. The model training method for the scenario of migration from Ethereum to Bitcoin is to swap the usage strategy of the Bitcoin and Ethereum datasets from the previous scenario. All methods are tested on the binary classification task of Ethereum malicious accounts. The results are shown in Table \ref{tab: Scalability-a}.

\begin{table*}[htb]  
	\tiny  
	\setlength{\abovecaptionskip}{0cm}  
	\setlength{\belowcaptionskip}{-0.2cm}  
	\caption{Scalability Comprison of Different Methods.}  
	\label{tab: Scalability-a}  
	\renewcommand\arraystretch{1.1}  
        \begin{threeparttable}
	\resizebox{\textwidth}{!}{  
		\begin{tabular}{clccccccccccc}  
			\hline  
			\multirow{2}{*}{Platform Compatibility} & \multirow{2}{*}{Metric} & \multicolumn{9}{c}{Method} \\ \cline{3-11}  
			& & CNN & ABGRL & DT & KNN & LR & MLP & FICL & BERT & \textbf{ShadowEyes} \\  
			\hline  
			\multirow{3}{*}{BTC to ETH}  
			& Pre($\%$) & 57.83 & 25.00 & 25.00 & 24.95 & 27.70 & 32.72 & 78.85 & 80.01 & \textbf{90.83} \textcolor{red}{\scalebox{0.7}{$\blacktriangle$ 10.82}} \\  
			& Recall($\%$) & 50.53 & 50.09 & 50.00 & 49.82 & 32.66 & 38.35 & 75.23 & 79.54 & \textbf{90.46} \textcolor{red}{\scalebox{0.7}{$\blacktriangle$ 10.92}} \\  
			& F1($\%$) & 53.98 & 33.39 & 33.33 & 33.23 & 29.98 & 35.33 & 77.00 & 79.77 & \textbf{90.64} \textcolor{red}{\scalebox{0.7}{$\blacktriangle$ 10.87}} \\  
			\hline  
			\multirow{3}{*}{ETH to BTC}  
			& Pre($\%$) & 35.47 & 35.84 & 66.96 & 43.97 & 41.31 & 38.73 & 69.53 & 79.21 & \textbf{88.73} \textcolor{red}{\scalebox{0.7}{$\blacktriangle$ 9.52}} \\  
			& Recall($\%$) & 40.32 & 39.78 & 50.63 & 74.00 & 64.34 & 48.18 & 72.52 & 76.91 & \textbf{87.35} \textcolor{red}{\scalebox{0.7}{$\blacktriangle$ 10.44}} \\  
			& F1($\%$) & 37.80 & 37.72 & 57.69 & 55.56 & 50.00 & 42.99 & 71.13 & 78.00 & \textbf{87.99} \textcolor{red}{\scalebox{0.7}{$\blacktriangle$ 9.99}} \\
			\hline
		\end{tabular}
	}
        \begin{tablenotes}
        \scriptsize       
        \item[1] We use \textcolor{red}{\scalebox{0.7}{$\blacktriangle$}} to highlight the improvement of classification performance compared with the SOTA BERT.
        \end{tablenotes}
        \end{threeparttable}
\end{table*}

Experimental results demonstrate that our ShadowEyes exhibits excellent across-platform performance. In scenarios involving migration between the Ethereum and Bitcoin platforms, ShadowEyes outperformed the state-of-the-art method, BERT, by nearly 10\% across all metrics. This superior performance can be attributed to ShadowEyes' integration of robust transaction representation methods, highly transferable contrastive learning, and effective data augmentation techniques. These features make it efficient for across-platform detection of malicious cryptocurrency transactions.

\begin{table*}[htb]    
    \tiny    
    \setlength{\abovecaptionskip}{0cm}    
    \setlength{\belowcaptionskip}{-0.2cm}    
    \caption{The Evaluation Results of Different Methods with Imbalanced Dataset.}    
    \label{tab: comparison}    
    \renewcommand\arraystretch{1.1}
    \begin{threeparttable}
    \resizebox{\textwidth}{!}{    
        \begin{tabular}{ccccccccccc}    
            \hline    
            \multirow{2}{*}{Metric}   & \multirow{2}{*}{Number of samples} & \multicolumn{9}{c}{Method}                                                              \\ \cline{3-11}     
            & &CNN &ABGRL & DT &KNN &LR &MLP& FICL &BERT&\textbf{ShadowEyes} \\     
            \hline    
    \multirow{5}{*}{Precision($\%$)}     
    &Ini. &83.78 &92.13 &86.28 &86.52 &79.76 &88.44 &78.32 &87.18 &\textbf{94.35}\textcolor{red}{\scalebox{0.7}{$\blacktriangle$ 7.17}}\\    
    &1:5 &79.90 &89.00 &86.94 &80.55 &72.40 &85.25 &78.01 &86.75 &\textbf{91.25}\textcolor{red}{\scalebox{0.7}{$\blacktriangle$ 4.50}}\\    
    &1:25&70.35 &80.10 &84.75 &75.25 &59.40 &79.60 &77.77 &84.66 &\textbf{89.63}\textcolor{red}{\scalebox{0.7}{$\blacktriangle$ 4.97}}\\    
    &1:50&66.40 &77.10 &84.95 &73.80 &53.60 &78.20 &75.54 &85.20 &\textbf{88.90}\textcolor{red}{\scalebox{0.7}{$\blacktriangle$ 3.70}}\\    
    &1:100&60.60 &60.95 &73.70 &60.80 &50.35 &67.80 &75.21 &79.87 &\textbf{83.90}\textcolor{red}{\scalebox{0.7}{$\blacktriangle$ 4.03}}\\    
    \hline    
    \multirow{5}{*}{Recall($\%$)}      
    &Ini. &80.41 &89.87 &83.45 &83.17 &67.11 &86.87 &70.27 &86.58 &\textbf{92.87}\textcolor{red}{\scalebox{0.7}{$\blacktriangle$ 6.29}}\\    
    &1:5 &72.52 &82.23 &80.86 &72.68 &65.09 &78.35 &71.95 &84.25 &\textbf{85.65}\textcolor{red}{\scalebox{0.7}{$\blacktriangle$ 1.40}}\\    
    &1:25&62.81 &71.56 &77.04 &66.95 &55.19 &71.17 &60.70 &78.75 &\textbf{84.05}\textcolor{red}{\scalebox{0.7}{$\blacktriangle$ 5.30}}\\    
    &1:50&59.81 &68.59 &77.07 &65.65 &51.86 &69.60 &54.25 &79.40 &\textbf{82.10}\textcolor{red}{\scalebox{0.7}{$\blacktriangle$ 2.70}}\\    
    &1:100&55.94 &56.15 &65.59 &56.05 &50.18 &60.85 &50.85 &66.70 &\textbf{75.58}\textcolor{red}{\scalebox{0.7}{$\blacktriangle$ 8.88}}\\    
    \hline    
    \multirow{5}{*}{F1($\%$)}      
    &Ini. &82.06 &90.99 &84.84 &84.81 &72.89 &87.65 &74.08 &86.88 &\textbf{93.60}\textcolor{red}{\scalebox{0.7}{$\blacktriangle$ 6.72}}\\    
    &1:5 &76.03 &85.48 &83.79 &76.41 &68.55 &81.65 &74.86 &85.48 &\textbf{88.36}\textcolor{red}{\scalebox{0.7}{$\blacktriangle$ 2.88}}\\    
    &1:25&66.36 &75.59 &80.71 &70.86 &57.22 &75.15 &68.18 &81.60 &\textbf{86.75}\textcolor{red}{\scalebox{0.7}{$\blacktriangle$ 5.15}}\\    
    &1:50&62.93 &72.59 &80.82 &69.49 &52.72 &73.65 &63.15 &82.20 &\textbf{85.36}\textcolor{red}{\scalebox{0.7}{$\blacktriangle$ 3.16}}\\    
    &1:100&58.17 &58.45 &69.41 &58.33 &50.26 &64.14 &60.68 &72.69 &\textbf{79.58}\textcolor{red}{\scalebox{0.7}{$\blacktriangle$ 6.89}}\\    
    \hline    
    \end{tabular}    
    }
    \begin{tablenotes}
        \scriptsize       
        \item[1] We use \textcolor{red}{\scalebox{0.7}{$\blacktriangle$}} to highlight the improvement of classification performance compared with the SOTA BERT.
        \end{tablenotes}
        \end{threeparttable}
        \vspace{-0.2cm}
\end{table*}

\subsection{Evaluation on Imbalanced Dataset}\label{subsec:Classification}
We evaluate the classification performance of ShadowEyes under different positive and negative sample ratios. Among them, $Ini.$ represents the model trained on the Bitcoin labeled address dataset mentioned in Section \ref{subsec:data}, while the others represent the model's performance when trained under conditions with malicious to normal account ratios of 1:5, 1:25, 1:50, and 1:100. Noted that since the Bitcoin labeled dataset contains 11 labels, the results under this condition reflect the model's multiclass performance, while the other four scenarios reflect the model's binary classification performance. According to the results shown in Tables \ref{tab: comparison}.

ShadowEyes achieves the best detection performance under each ratio setting. Even in highly imbalanced dataset scenario (1:100), ShadowEyes still has an F1 score of 79.58\%, which is 6.59\% higher than BERT. The F1 score of FICL is consistently more than 10\% lower than BERT under each ratio setting. The primary difference between FICL and ShadowEyes is their data augmentation strategies, suggesting that our data augmentation approach allows the model to better capture diverse malicious transaction patterns, thereby improving classification performance.

ShadowEyes has the relatively stable detection performance under each ratio setting. When ratio setting enlarges from 1:5 to 1:50, the F1 score of ShadowEyes reduces by 14.02\%, while other machine learning methods without pre-training all dropped by more than 20\%, with the ABGRL method even dropping by 32.54\%. This is because TxGraph can capture more discriminative features between different malicious transactions than other representations. It allows ShadowEyes to accurately identify malicious transaction, even in the highly imbalanced dataset scenario.

\subsection{Evaluation on Few-shot Learning}
We further evaluate the performance of ShadowEyes in scenarios characterized by limited labeled training data. Specifically, we evaluate the model's performance under six different training sizes: 10, 20, 50, 100, 200, and 500 samples. The results are shown in Table \ref{tab: small-sample }.

ShadowEyes achieves the best detection performance across all dataset sizes. In the scenario with only 10 training samples, ShadowEyes' F1 score was 78.22\%, roughly on par with BERT. However, as the sample size increased, the F1 score of ShadowEyes rose at a relatively higher rate. When the sample size reached 500, ShadowEyes' F1 score increased to 91.76\%, with an increase of 13.54\%. This improvement is due to ShadowEyes' effective pre-training, which allows it to initially classify positive and negative samples, and the increased number of labeled datasets further enhances its classification ability.

Note that among the methods requiring pre-training, the F1 score of FICL is consistently more than 10\% lower than BERT under each size setting. The primary difference between FICL and ShadowEyes is their data augmentation strategies, suggesting that our data augmentation approach allows the model to better capture diverse malicious transaction patterns, thereby improving classification performance.

\begin{table*}[htb]    
    \tiny    
    \setlength{\abovecaptionskip}{0cm}    
    \setlength{\belowcaptionskip}{-0.2cm}    
    \caption{Performance Comparison in Few-shot Scenario.}    
    \label{tab: small-sample }       
    \renewcommand\arraystretch{1.1}
    \begin{threeparttable}
    \resizebox{\textwidth}{!}{    
        \begin{tabular}{ccccccccccccc}    
            \hline    
            \multirow{2}{*}{Metric}   & \multirow{2}{*}{Number of samples} & \multicolumn{9}{c}{Method(\%)}                                                              \\ \cline{3-11}     
            & & CNN & ABGRL & DT & KNN & LR & MLP & FICL & BERT & \textbf{ShadowEyes} \\     
            \hline    
    \multirow{6}{*}{Precision(\%)}      
            & 10 & 59.90 & 77.50 & 76.45 & 78.45 & 74.55 & 74.10 & 68.65 & 78.43 & \textbf{78.45} \textcolor{red}{\scalebox{0.7}{$\blacktriangle$ 0.02}}\\    
            & 20 & 74.90 & 79.75 & 77.60 & 79.05 & 78.00 & 79.70 & 74.33 & 79.15 & \textbf{80.10} \textcolor{red}{\scalebox{0.7}{$\blacktriangle$ 0.95}}\\    
            & 50 & 77.85 & 82.85 & 79.25 & 81.80 & 79.20 & 82.75 & 63.59 & 83.21 & \textbf{84.15} \textcolor{red}{\scalebox{0.7}{$\blacktriangle$ 0.94}}\\    
            & 100 & 78.40 & 85.75 & 81.90 & 84.00 & 78.40 & 83.60 & 78.13 & 84.86 & \textbf{87.65} \textcolor{red}{\scalebox{0.7}{$\blacktriangle$ 2.79}}\\    
            & 200 & 83.05 & 88.10 & 82.55 & 84.50 & 78.20 & 85.80 & 79.45 & 86.56 & \textbf{88.45} \textcolor{red}{\scalebox{0.7}{$\blacktriangle$ 1.89}}\\    
            & 500 & 80.60 & 89.40 & 86.00 & 87.15 & 79.20 & 85.75 & 80.08 & 87.97 & \textbf{89.90} \textcolor{red}{\scalebox{0.7}{$\blacktriangle$ 1.93}}\\    
            \hline    
    \multirow{6}{*}{Recall(\%)}      
            & 10 & 71.80 & 74.00 & 72.50 & 71.00 & 71.90 & 69.80 & 65.90 & 77.40 & \textbf{78.00} \textcolor{red}{\scalebox{0.7}{$\blacktriangle$ 0.60}}\\    
            & 20 & 74.80 & 79.60 & 81.90 & 77.70 & 80.90 & 77.40 & 74.30 & 80.10 & \textbf{84.00} \textcolor{red}{\scalebox{0.7}{$\blacktriangle$ 3.90}}\\    
            & 50 & 77.30 & 81.40 & 83.20 & 80.70 & 80.40 & 81.20 & 62.00 & 84.20 & \textbf{85.00} \textcolor{red}{\scalebox{0.7}{$\blacktriangle$ 0.80}}\\    
            & 100 & 83.80 & 79.40 & 87.80 & 83.70 & 80.30 & 83.70 & 77.95 & 85.60 & \textbf{88.00} \textcolor{red}{\scalebox{0.7}{$\blacktriangle$ 2.40}}\\    
            & 200 & 80.60 & 81.60 & 82.00 & 75.40 & 80.80 & 86.10 & 76.24 & 87.20 & \textbf{91.60} \textcolor{red}{\scalebox{0.7}{$\blacktriangle$ 4.40}}\\    
            & 500 & 85.00 & 85.80 & 87.40 & 82.10 & 80.50 & 85.20 & 79.65 & 89.20 & \textbf{93.70} \textcolor{red}{\scalebox{0.7}{$\blacktriangle$ 4.50}}\\    
            \hline    
    \multirow{6}{*}{F1(\%)}      
            & 10 & 65.31 & 75.71 & 74.42 & 74.54 & 73.20 & 71.89 & 67.25 & 77.91 & \textbf{78.22} \textcolor{red}{\scalebox{0.7}{$\blacktriangle$ 0.31}}\\    
            & 20 & 74.85 & 79.67 & 79.69 & 78.37 & 79.42 & 78.53 & 74.31 & 79.62 & \textbf{82.00} \textcolor{red}{\scalebox{0.7}{$\blacktriangle$ 2.38}}\\    
            & 50 & 77.57 & 82.12 & 81.18 & 81.25 & 79.80 & 81.97 & 62.78 & 83.70 & \textbf{84.57} \textcolor{red}{\scalebox{0.7}{$\blacktriangle$ 0.87}}\\    
            & 100 & 81.01 & 82.45 & 84.75 & 83.85 & 79.34 & 83.65 & 78.04 & 85.23 & \textbf{87.82} \textcolor{red}{\scalebox{0.7}{$\blacktriangle$ 2.59}}\\    
            & 200 & 81.81 & 84.73 & 82.27 & 79.69 & 79.48 & 85.95 & 77.81 & 86.88 & \textbf{90.00} \textcolor{red}{\scalebox{0.7}{$\blacktriangle$ 3.12}}\\    
            & 500 & 82.74 & 87.56 & 86.69 & 84.55 & 79.84 & 85.47 & 79.86 & 88.58 & \textbf{91.76} \textcolor{red}{\scalebox{0.7}{$\blacktriangle$ 3.18}}\\    
            \hline    
        \end{tabular}    
    } 
    \begin{tablenotes}
        \scriptsize       
        \item[1] We use \textcolor{red}{\scalebox{0.7}{$\blacktriangle$}} to highlight the improvement of classification performance compared with the SOTA BERT.
        \end{tablenotes}
        \end{threeparttable}
        \vspace{-0.2cm}
\end{table*}

Overall, ShadowEyes not only performs well initially in small sample scenarios but also shows faster and greater improvement in recognition performance as the sample size increases compared to other methods.

\subsection{Ablation Experiment}
We further evaluate the effectiveness of the three main module that make up ShadowEyes, including transaction representation, the contrastive learning encoder.

We evaluate the effectiveness of each component of ShadowEyes in different scenarios. We define w/o Repre. and w/o CL-Encoder as methods to remove transaction representation and contrastive learning encoder, respectively, and compare them with the complete ShadowEyes method to evaluate the effectiveness of these two modules. 

The experimental results are presented in Table \ref{tab: Ablation}, highlight the impact of different components on model performance. When the transaction representation module is removed (i.e., w/o Repre.), there is an overall performance decline of approximately 10\%. This underscores the significant enhancement provided by the address behavior representation module to the ShadowEyes model. Additionally, the removal of the contrastive learning encoder (i.e., w/o CL-Encoder) results in a substantial drop in the F1 score across all scenarios, averaging around 60\%, which represents a decrease of about 30\% compared to the full ShadowEyes. 

\begin{table}[htbp]
\centering
\caption{Module Effectiveness Evaluation}
\label{tab: Ablation}
\renewcommand\arraystretch{1.1}
\resizebox{0.95\linewidth}{!}{
\begin{tabular}{ccccc}
\hline
Method & Metric & Normal & Small Sample & Imbalanced \\ \hline
\multirow{3}{*}{w/o Repre.} 
    & Pre & 82.96\% & 68.40\% & 76.85\% \\
    & Recall & 80.23\% & 75.60\% & 68.61\% \\
    & F1 & 81.57\% & 71.82\% & 72.49\% \\ \hline
\multirow{3}{*}{w/o CL-Encoder} 
    & Pre & 66.55\% & 50.65\% & 66.50\% \\
    & Recall & 61.61\% & 60.00\% & 60.15\% \\
    & F1 & 63.98\% & 54.93\% & 63.16\% \\ \hline
\multirow{3}{*}{ShadowEyes} 
    & Pre & 94.35\% & 78.45\% & 88.90\% \\
    & Recall & 92.87\% & 78.00\% & 82.10\% \\
    & F1 & \textbf{93.60\%} & \textbf{78.22\%} & \textbf{85.36\%} \\ \hline
\end{tabular}
}
\end{table}

To further evaluate the classification effectiveness of the pre-training part of ShadowEyes on positive and negative samples, we utilized the t-SNE algorithm to visualize ShadowEyes and ShadowEyes (w/o CL-Encoder), where different colors represent different categories. Compared to ShadowEyes (w/o CL-Encoder) that only uses prediction loss $\mathcal{L}_f$, ShadowEyes guided by contrastive loss achieves more distinct inter-class separation and intra-class compactness. This demonstrates the effectiveness of pre-training on unlabeled data in learning behavioral pattern differences.

\begin{figure}[htbp]
    \centering
    \subfigure[Before pre-training]{
        \includegraphics[width=0.44\linewidth]{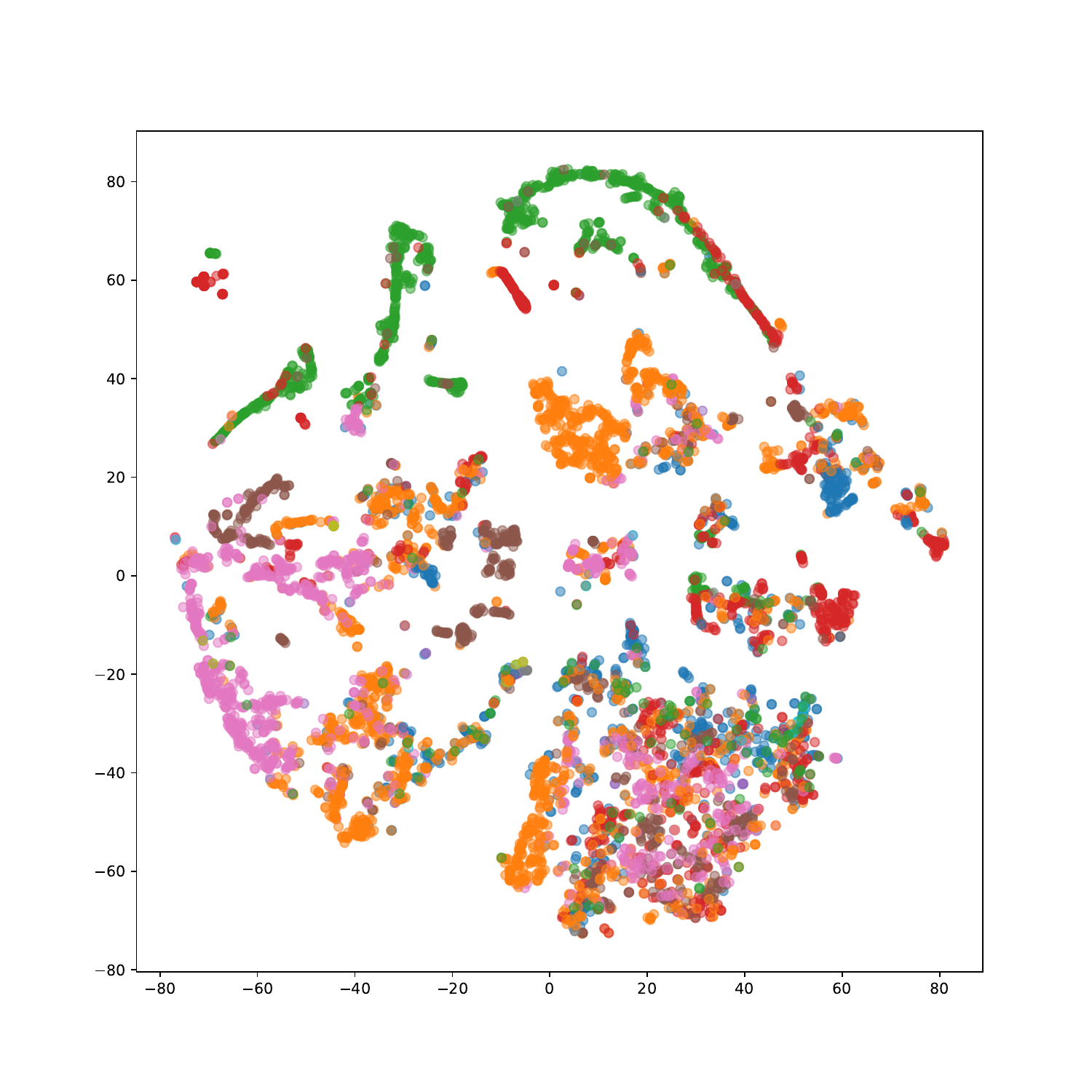}
    }
    \hfill
    \subfigure[After pre-training]{
        \includegraphics[width=0.44\linewidth]{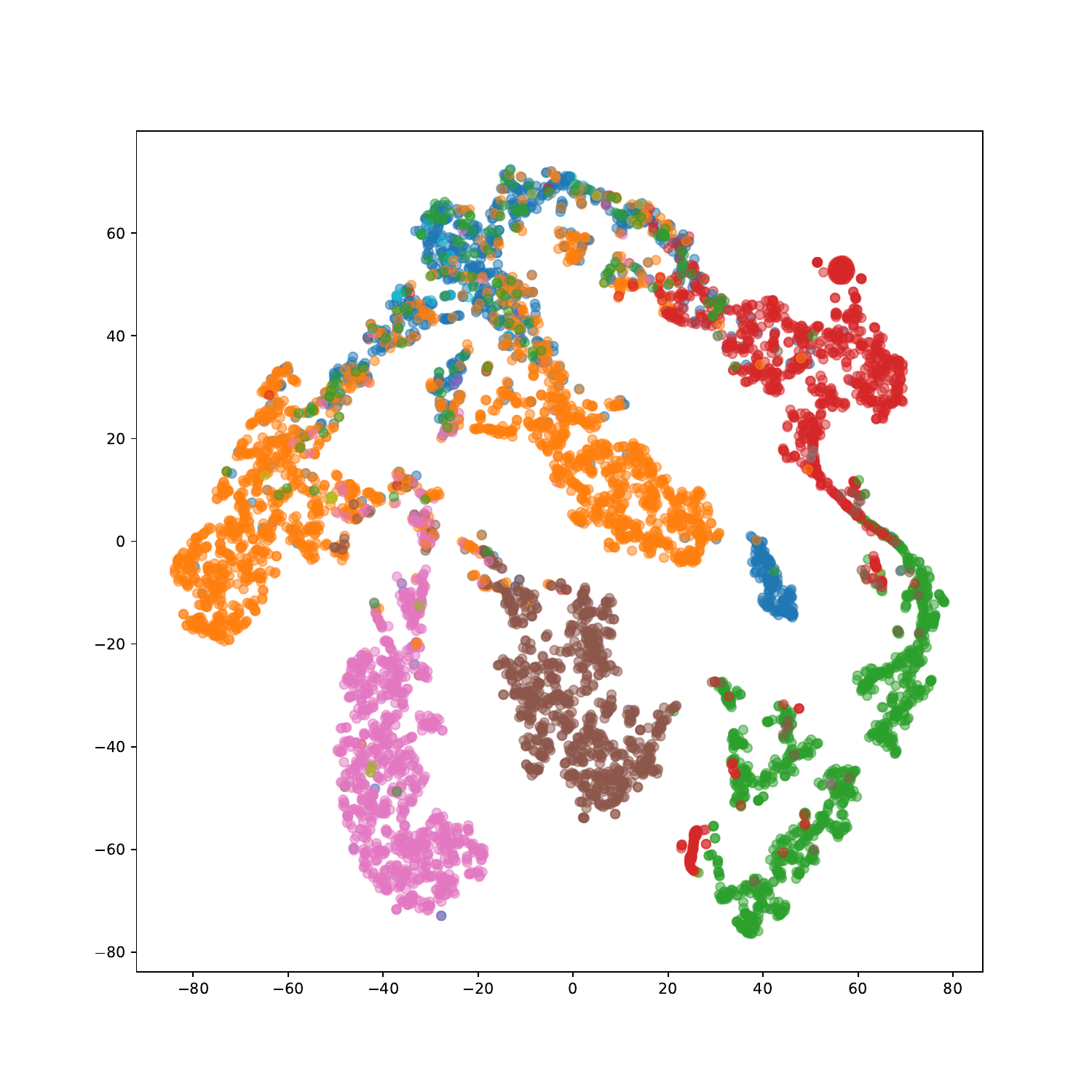}
    }
    \caption{The t-SNE visualization of the sample embeddings learned by ShadowEyes with and without pre-traning}
    \label{fig:t-SNE}
\end{figure}

In summary, this section verifies the effectiveness of each component of ShadowEyes and also demonstrates the positive impact on improving address classification accuracy when these modules work together.

\section{Discussion}\label{sec:discussion}
In this section, we discuss the limitations of ShadowEyes and potential directions for enhancing our methods.

\textbf{Training Overhead}. The pre-training time of the contrastive learning module is relatively long, as any two samples in the data set will be paired for contrastive learning. Moreover, in the contrastive learning module, we need a sufficient number of samples for training to fully learn the transaction feature extraction method, which also increases the training time. However, this does not affect the application in practical scenarios, as unlabeled transaction data is easy to obtain, giving us ample time to train the model before actual deployment. In future work, we consider using large language models like GPT-4, which have been pre-trained on extensive text data, as our pre-training model. Fine-tuning on this foundation can reduce our total training time.

\textbf{Generalization}. In this paper, constrained by publicly available cryptocurrency datasets, we evaluated ShadowEyes only on the Bitcoin dataset BABD-13 \cite{2024TIFS-Dataset} and the Ethereum dataset \cite{ETH-Dataset}. However, besides Bitcoin and Ethereum, there are other major cryptocurrency platforms with high market value, such as Ripple, Binance, and TRON. In future work, we plan to collect and construct malicious transaction datasets from these platforms and conduct further evaluations of ShadowEyes on these datasets.

\section{Conclusion}\label{sec:Conclusion}
Malicious transaction detection has become crucial for safeguarding the healthy development of blockchain ecosystems. This paper proposes ShadowEyes, a novel cryptocurrency malicious transaction detection method based on graph contrastive learning. The node interaction features of each malicious account and its neighboring nodes are extracted as transaction representations. Data augmentation techniques, such as transaction amount splitting and adding random time delays, are proposed to generate enhanced samples, which are used for constructing positive pairs. Based on these pairs, using only unlabeled transaction samples, a feature extraction model can be trained. Leveraging just a few labeled transaction samples, the model can achieve effective malicious transaction detection. Experiments are carried out to demonstrate ShadowEyes's effectiveness in detecting unknown malicious transactions and across-platform malicious transactions. In future work, we plan to gather and incorporate labeled transactions from additional platforms such as Ripple, Binance, and TRON. This expansion will enhance ShadowEyes's ability to detect malicious transactions across a broader range of platforms.

\if CLASSOPTIONcaptionsoff
\newpage
\fi



\bibliographystyle{IEEEtran}
\bibliography{refs}

\end{document}